\documentclass[%
 manuscript,%
 amssymb, amsmath,%
 aip,cha,%
]{revtex4-1}
\usepackage{float}
\usepackage{docs}%
\usepackage{bm}%
\usepackage{graphicx}
\usepackage{epstopdf}
\usepackage{longtable}
\expandafter\ifx\csname package@font\endcsname\relax\else
 \expandafter\expandafter
 \expandafter\usepackage
 \expandafter\expandafter
 \expandafter{\csname package@font\endcsname}%
\fi
\begin{document}

\title{An Environment-dependent Semi-Empirical Tight Binding Model Suitable for Electron Transport in Bulk Metals, Metal Alloys, Metallic Interfaces and Metallic Nanostructures I - Model and Validation}%

\author{Ganesh Hegde}%
\email{ghegde@purdue.edu}
\affiliation{Network for Computational Nanotechnology (NCN),\\Department of Electrical and Computer Engineering, Purdue University\\West Lafayette, Indiana 47907, USA}%
\author{Michael Povolotskyi}
\affiliation{Network for Computational Nanotechnology (NCN),\\Department of Electrical and Computer Engineering, Purdue University\\West Lafayette, Indiana 47907, USA}%
\author{Tillmann Kubis}
\affiliation{Network for Computational Nanotechnology (NCN),\\Department of Electrical and Computer Engineering, Purdue University\\West Lafayette, Indiana 47907, USA}%
\author{Timothy Boykin}
\affiliation{Department of Electrical and Computer Engineering, University of Alabama,\\Huntsville, Alabama, USA}%
\author{Gerhard Klimeck}
\email{gekco@purdue.edu}
\affiliation{Network for Computational Nanotechnology (NCN),\\Department of Electrical and Computer Engineering, Purdue University\\West Lafayette, Indiana 47907, USA}%

\date{\today}%

\begin{abstract}
Semi-Empirical Tight Binding (TB) is known to be a scalable and accurate atomistic representation for electron transport for realistically extended nano-scaled semiconductor devices that might contain millions of atoms. In this paper an environment-aware and transferable TB model suitable for electronic structure and transport simulations in technologically relevant metals, metallic alloys, metal nanostructures and metallic interface systems is described. Part I of this paper describes the development and validation of the new TB model.

The new model incorporates intra-atomic diagonal and off-diagonal elements for implicit self-consistency and greater transferability across bonding environments. The dependence of the on-site energies on strain has been obtained by appealing to the Moments Theorem that links closed electron paths in the system to energy moments of angular momentum resolved local density of states obtained \textit{ab} \textit{initio}. The model matches self-consistent DFT electronic structure results for bulk FCC metals with and without strain, metallic alloys, metallic interfaces and metallic nanostructures with high accuracy and can be used in predictive electronic structure and transport problems in metallic systems at realistically extended length scales.
\end{abstract}

\maketitle


\section{Introduction}

The degradation of metallic conductance due to quantum confinement, interface and surface roughness scattering and grain-boundary scattering is a well-known technological problem of current relevance. The importance of this problem can be evidenced from the fact that successive editions of the International Technology Roadmap for Semiconductors \cite{itrs} have marked the problem as one with no known or foreseen mitigative solutions. 

Semiclassical models of conductance degradation in metals are largely based on the seminal work of Fuchs and Sondheimer (F-S) on surface scattering \cite{sondheimer1952mean, fuchs1938conductivity}, and Mayadas and Shatzkes (M-S) on grain boundary scattering \cite{mayadas1970electrical}. A feature common to early and recent variants of these models is the use of the Boltzmann transport equation and averaged empirical parameters such as specularity, average reflectivity, and average grain size. These parameters are fit \textit{a} \textit{posteriori} to experimental conductivity data to relate bulk (ideal) and degraded transport parameters such as conductance and conductivity. Sambles \cite{sambles1983resistivity} and Josell et al. \cite{josell2009size} provide a detailed description of such semiclassical variants.

The most serious drawback in such \textit{a} \textit{posteriori} models is the lack of predictive capability from a materials design perspective. Atomistic information such as strain, interface bonding, alloying and surface roughness are expected to affect conductance significantly.  It is not possible to predict how these atomistic effects affect conductance unless experimental data is available to fit empirical models that use averaged parameters or Boltzmann transport. It is especially disconcerting that conflicting insights may result from the use of model variants on similar experimental data. For instance, Steinh\"{o}gl et al.\cite{steinhogl2002size} \cite{steinlesberger2002electrical} report a semiclassical variant of F-S and M-S that shows grain-boundary scattering dominating for wire thickness in the sub-50 nm regime. This conclusion conflicts with the semiclassical model of Graham et al. \cite{graham2010resistivity}, who report a conductivity degradation that is dominated by surface and line-edge roughness scattering in the sub-50 nm regime with excellent fits to experimental data.

Atomistic quantum mechanical modeling of electronic transport can provide accurate, unambiguous insight into the dependence of metallic conductance on a variety of controlling factors - strain, alloying, dimensionality, interface bonding and grain boundaries. Recent \textit{ab} \textit{initio} DFT based studies have provided important insights into electronic transport in metallic systems \cite{PhysRevB.81.045406, zhou2008resistance, kharche2011comparative, feldman2010simulation, PhysRevB.79.155406}. Metallic systems such as polycrystalline interconnects, however, contain grains having several thousand to a few million atoms. In general DFT methods that deliver the needed accuracy for electronic structure details and carrier transport near equilibrium are limited to a maximum system size of upto thousands of atoms.

Semi-Empirical Tight Binding (TB), when properly parameterized, is known to be a scalable and accurate atomistic representation for electron transport for realistically extended nano-scaled semiconductor devices that might contain millions of atoms. TB electronic structure calculations in multi-million atom semiconductor systems such as quantum dots \cite{klimeck2002development, usman2009moving}, single-impurity devices  \cite{lansbergen2008gate}, disordered SiGe quantum wells \cite{kharche2007valley} have been successfully used to quantitatively explain and predict experimental results.

In the domain of semiconductor-based nanoelectronic transport research the TB basis sets have established themselves in conjunction with the NEGF (Non-Equlibrium Green's Function) formalism as the accepted state-of-the-art for quantitative device design \cite{bowen1997quantitative, steiger2011nanoelectronic}. 

Crucial to the success of TB in enabling realistically extended electronic structure and carrier transport simulations are the following salient features.
\begin{enumerate}
\item \textit{Nearest Neighbor (NN) or second Nearest Neighbor (2NN) couplings} - this enables system representation with a Hamiltonian matrix having significant sparsity and significantly reduces computational demands.
\item \textit{Parameterized Hamiltonian} - This obviates the need to compute multi-center integrals in evaluating Hamiltonian matrix elements.
\item \textit{Orthogonality of the basis} - This enables a sparse system matrix representation and simplified numerical approaches. It also avoids computation and inversion of a non-unity overlap matrix and significantly improves computational efficiency in electronic transport problems.
\item \textit{Transferability} within classes of material systems such as InGaAlAs or SiGe.
\item \textit{Atomistic representation} of strain and realistic distortion.
\item \textit{Locality of the basis} to represent finite nano-structures with finite contact regions.
\end{enumerate}

The situation for TB models of metals, however, is quite different. Existing TB models of metals in the literature have several important shortcomings that prevent them from being applied to predictive, scalable electronic structure and transport calculations on realistic metallic systems of interest. Early TB models \cite{papaconstantopoulos1986handbook} of metals were parameterized for bulk interactions and are not applicable to systems involving strain, alloying, interfaces and non bulk-like atomic coordination.  The most cited environment-dependent TB model for metals is the so-called NRL-TB model \cite{cohen1994tight, mehl1996applications}. The model resulted in environment-dependent TB parameterizations in a large number of elemental metals. This method was originally developed for total energy applications such as molecular dynamics and phonon calculations in metals. It uses a non-orthogonal basis, contains up to 300 independent parameters and incorporates interactions of up to 135 neighbors. Additionally, when applied to low-dimensional systems such as thin films, it has unphysical charge transfer on surface atoms \cite {PhysRevB.58.9721, xie2001transferable}. 

NRL-TB variants by Barreteau et al. \cite{PhysRevB.58.9721} and Xie et al. \cite{xie2001transferable} \cite{xie2001tight} address issues of self-consistency and non-orthogonality present in the NRL-TB model. Both models, however, still retain interactions of up to fourth or fifth nearest neighbors. Even the absence of an applied potential, both variants require iterations to self-consistency in order to correctly represent charge transfer effects in non bulk-like bonding environments. This leads to a dramatic loss of computational efficiency. NRL-TB and its variants have been applied to systems such as atomic clusters and surfaces. It is unclear if these models are suitable for inhomogeneously strained systems such as grain boundaries, hetero-metal interfaces and metallic alloys. 

Importantly, most TB parameterizations of metals and non-metals (with one notable exception \cite{boykin2010strain}) use \textit{ad} \textit{hoc} models for the variation of TB on-site elements with changes in the environment. This arbitrariness leads to a lack of physical transparency in TB models and increases the number of independent parameters that need to be used.

In this paper, a new environment-aware TB model suitable for scalable electronic transport in FCC metals is reported. The model includes intra-atomic diagonal and off-diagonal elements for improved transferability and implicit self-consistency. The variation of on-site energies with strain is obtained by appealing to the Moments Theorem \cite{sutton1993electronic, gaspard1973density} that links closed electron paths in the crystal to energy moments of the Local Density of States (LDOS). The new model is orthogonal and includes interactions only up to first nearest neighbor (1NN) for bulk FCC metals and second nearest neighbor (2NN) for non bulk-like coordinations. Though the model is applied in this series of papers to FCC metals Cu, Au, Ag and Al, the model is general enough for application to semiconductors and metals with other ground state crystal structures.

The rest of this paper is organized as follows. First, the formalism of the present model is presented, followed by details of the fitting procedure used to obtain TB parameters. The model is then validated by comparing it to DFT calculations of several metallic systems. Owing to the large number of results generated for various metals, only representative results are included in the main body of the paper. The remaining results and the optimized TB parameters are included in the appendix. Finally, key features of the model are summarized. Applications of the new model to relevant problems of interest are deferred to part II of this series of papers.

\section{Model}
In the interest of brevity, it is assumed that the reader is familiar with the Slater-Koster two-center TB formalism \cite{slater1954simplified}. For an excellent review of TB and the variety of TB models that exist in the literature, the interested reader is referred to the review by Goringe et al. \cite{goringe1997tight} 

\subsection{Intra-atomic diagonal and off-diagonal elements in the new model}
In the usual Slater Koster (SK) two-center TB formalism \cite{slater1954simplified}, a change in atom environment due to strain, alloying, interfaces or a change in coordination changes only the off-diagonal inter-atomic block of the Hamiltonian. Since the on-site block is usually diagonal and contains only on-site energies, changes to atom potential due to a change in environment are then included by adding correction terms \cite{goringe1997tight, xie2001transferable, PhysRevB.58.9721, boykin2002diagonal} to the diagonal block and/or iterating to self-consistency. 

In our new TB model, intra-atomic, diagonal and off-diagonal elements are included in the on-site block of the TB Hamiltonian in addition to on-site energies. The intra-atomic elements $I_{ll'm}$ (where $l$ and $l'$ are angular momenta $s$, $p$ or $d$ of orbitals on the same atom and $m$ is $\sigma$, $\pi$ or $\delta$) incorporate the effects of neighbor potentials as well as L\"{o}wdin renormalization \cite{boykin2002diagonal, lowdin1950non} that maintains an orthonormal basis under displacement of neighbor atoms. They are analogous to two-center Slater-Koster inter-atomic elements $V_{ll'm}$.

It is important to realize that because a significant contribution to the intra-atomic integrals comes from the region nearest to the common nucleus, the $I_{ll'm}$ will not necessarily follow the expected signs of the corresponding $V_{ll'm}$. Also, because the two atoms in $I_{ll'm}$ have a common center, there is no sign change under the exchange $ l \leftrightarrow l'$. 

As a specific example of an intra-atomic Hamiltonian matrix element in terms of the intra-atomic integrals, the $s-p_{z}$ term for orbitals centered on atomic site $i$ having $N$ neighbors within its bonding radius at locations indexed by $j$ is 

\begin{equation}
\langle s,i\mid \hat{H} \mid p_{z},i \rangle = \sum\limits_{j\in N}n_{ij}I_{sp\sigma}(\mathbf{R}_{ij})
\end{equation}
where $R_{ij}$ is the distance between atom $i$ and neighbor $j$ that belongs to the set of $N$ neighbors bonded to $i$ and $n_{ij}$ is the corresponding $z$-direction cosine.
  
Including intra-atomic diagonal and off-diagonal elements has desirable advantages from a physical perspective. It ensures that any change in the neighboring environment of an atom implicitly enters the on-site block of the TB Hamiltonian. Making the intra-atomic elements dependent on neighbor type and distance ensures that strain and disorder in metallic systems is not treated as a correction but is an integral part of the TB Hamiltonian. 

Additionally, effects such as crystal field splitting of on-site energies are also implicitly treated due to the inclusion of intra-atomic elements. By carefully parameterizing across a variety of neighbor environments, the present TB model ensures improved transferability and environment-awareness without explicitly iterating for self-consistency. 

From an implementation perspective, the intra-atomic block can be constructed similar to the off-diagonal inter-atomic block with the caveat that sign of the integrals does not change under interchange $l \leftrightarrow l'$ as mentioned above.

\subsection{Variation of TB parameters with strain}

\subsubsection{Two-Center Inter-atomic Integrals}
Most empirical TB models in the literature use a generalized version of Harrison's rule \cite{harrison1999elementary,harrison2012electronic, goringe1997tight} for two-center, off-diagonal,  inter-atomic integrals to describe their variation with interatomic distance. 
\begin{equation}
V_{ll'm} = {V^{(0)}}_{ll'm}({R^{(0)}}/R)^{\eta}
\end{equation}
where ${V^{(0)}}_{ll'm}$ represents the unstrained two-center elements, $R$ is the strained bond length between atoms and the superscript $(0)$ is used to indicate unstrained values. $\eta$ is the decay exponent, where $\eta=2$ represents the original proposal of Harrison. $\eta$ has also been used as an explicit fitting parameter in previous TB investigations \cite{boykin2002diagonal}.

Using such a functional form has been justified using physical arguments in the Linear Muffin Tin Orbital (LMTO)-TB theory of Andersen and Jepsen \cite{andersen1984explicit}. A similar approach is adopted in our model. The functional form used in our model is analogous to the generalized version of Harrison's law indicated above. To facilitate comparison, the functional form chosen here follows from existing TB models for metals in the literature \cite{PhysRevB.58.9721, xie2001tight, xie2001transferable, mehl1996applications}.
\begin{equation}
V_{ll'm} = {V^{(0)}}_{ll'm}\exp[-q_{ll'm}(\frac{R_{ij}}{{{R_{ij}}^{(0)}}}-1)]
\end{equation}
where $q$ is the decay exponent.

\subsubsection{On-Site terms}

Existing empirical TB models of metals and non-metals (with few notable exceptions \cite{boykin2010strain}) incorporate a wide variety of $ad$ $hoc$ functional forms to describe the variation of on-site energies with interatomic distance \cite{goringe1997tight}. A polynomial in the interatomic distance $R$ is usually formed and the coefficients of individual monomials are treated as fitting parameters. In the absence of a physical justification, a wide variation in the functional form of these polynomials and the number of fitting parameters is seen in the literature. For instance, the NRL-TB model has 4 independent parameters to describe on-site energies per angular momentum. The model of Mercer and Chou contains up to 8 parameters, while the model of Tang et al. contains 7 parameters per on-site term.

The general trend seen in these models is to use a large number of monomials in the polynomial in the hope that the variation of on-sites is captured accurately. While such a situation is tractable for elementary solids, it quickly becomes unmanageable in multi-atomic systems such as binary and ternary alloys or hetero-metal interfaces. The use of an arbitrary number parameters without a physical justification is undesirable. It results in a model that lacks physical transparency and makes the fitting process aribtrary and cumbersome.

The present model approaches this issue from a completely different perspective than has been used heretofore in the literature. Before describing the present model for the variation of on-site elements, it is essential to review the Moments Theorem of Cyrot-Lackmann \cite{gaspard1973density,sutton1993electronic}. 

The Moments Theorem links the Local Density Of States (LDOS) at a given atomic site to its neighboring environment. The theorem proves that closed electron paths of length $n$ hops beginning and ending at an atomic site $i$ correspond to the $n$th energy moment of the LDOS at $i$. Each electron hop corresponds to a distinct TB Hamiltonian matrix element. In the traditional application of this theorem, it is first assumed that Hamiltonian matrix elements are known. Closed electron paths of length $n$ hops are then computed using the TB Hamiltonian and the moments of the LDOS are written down in this theorem as follows.
\begin{equation}
\label{eq:moments}
{\mu_{i}}^{(n)} = \langle \phi_{i}\mid H^{n} \mid\phi_{i} \rangle
\end{equation}
where ${\mu_{i}}^{(n)}$ represents the $n$th energy moment of the LDOS at $i$ and $\phi_{i}$ is a linear combination of basis functions at site $i$. Thus, by simply knowing the TB Hamiltonian matrix elements, information about the LDOS in the form of its moments is obtained. The LDOS is then reconstructed from its moments without explicitly solving the Schr\"{o}dinger equation. The Moments Theorem has found wide application in the study of electronic structure in inherently non-periodic systems such as vacancies, impurities, and interfaces \cite{sutton1993electronic}.

In the present TB model, the Moments Theorem is used in reverse to obtain a trend in the variation of on-site energies with strain. First, the theorem for individual angular momenta is written following equation \ref{eq:moments} as follows
\begin{equation}
{\mu_{il}}^{(n)} = \langle \phi_{il}\mid H^{n} \mid\phi_{il} \rangle
\end{equation}
where $l$ is the angular momentum and $\phi_{il}$ represent $s$, $p$ or $d$ orbitals at site $i$. The quantity ${\mu_{il}}^{(n)}$ now represents the $n$th moment of the Momentum Resolved LDOS (MRLDOS) at atom $i$. 

From the Moments Theorem, the 1st Moment of MRLDOS corresponds to closed electron paths of length 1 hop that begin and end at the same orbital $\phi_{il}$. This is written as follows
\begin{equation}
\label{eq:firstmoment}
{\mu_{il}}^{(1)} = \langle \phi_{il}\mid H \mid\phi_{il} \rangle
\end{equation}

It is evident that the right hand side of equation \ref{eq:firstmoment} represents TB on-site elements. In the usual application of this theorem, once TB on-site elements are known, the first moments can be obtained directly from them. If the situation is reversed, as in the present situation, where the moments can be obtained from MRLDOS data given by DFT, the same equation can now be used to obtain unknown TB on-site energies. The moments of the MRLDOS at atomic site $i$ can be obtained as follows, 
\begin{equation}
{\mu_{il}}^{(n)} = \frac{\int{E^{n}g_{il}(E)dE}}{\int{g_{il}(E)dE}}
\end{equation}
where $g_{il}(E)$ is the MRLDOS at site $i$ obtained from DFT. Once this information is available equation \ref{eq:firstmoment} readily gives on-site elements. 

\begin{figure}[H]
	\centering
		\includegraphics[width=3.5in]{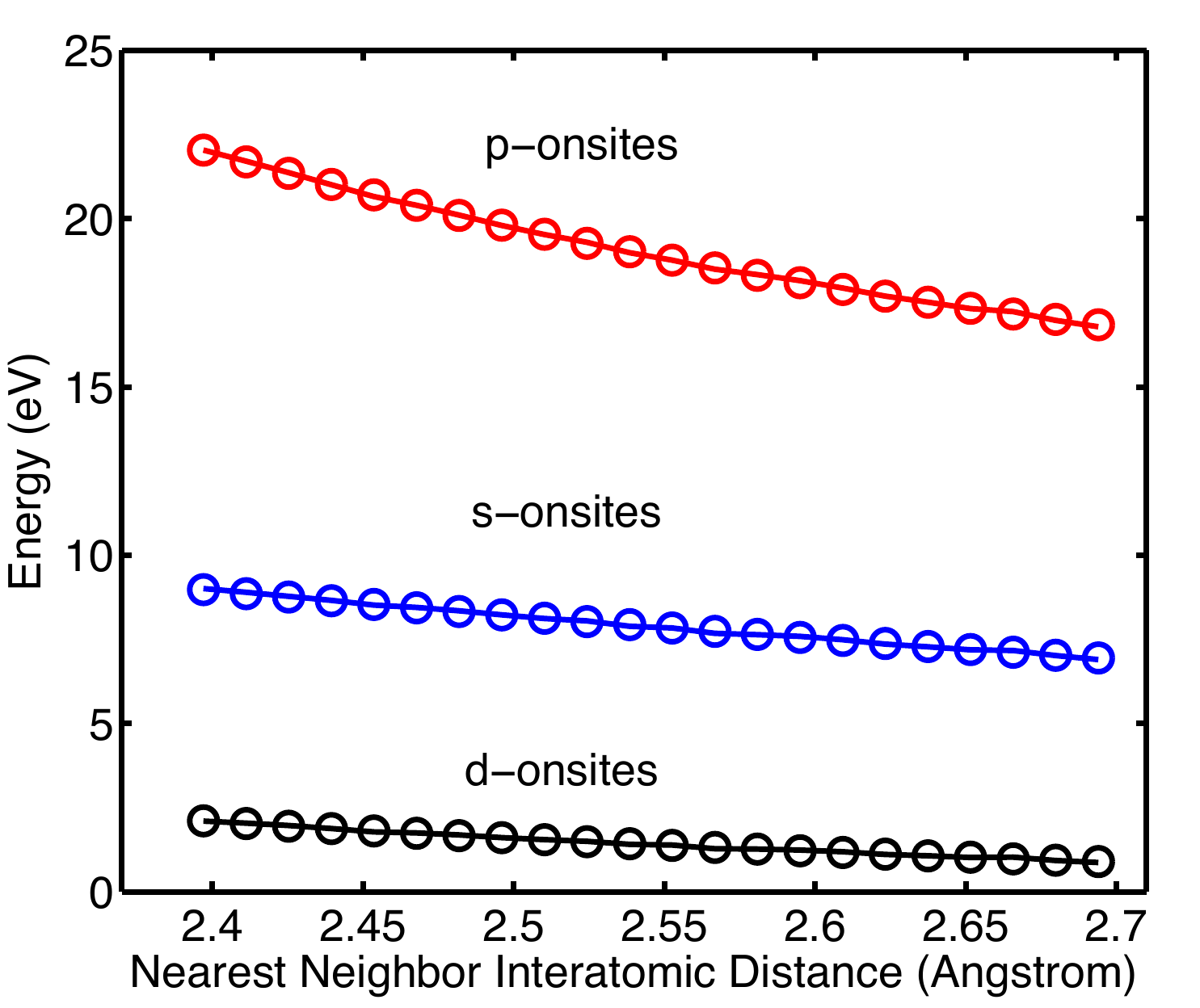}
	\caption{First moments of the $s$,$p$ and $d$ LDOS for bulk FCC Cu obtained from DFT versus interatomic distance indicated by solid lines. The circles are fits to the data using expression in equation 8}	
	\label{fig:On_site_fits_exponential}
\end{figure}

To compute the variation of on-site elements with strain, the first moments are computed from DFT MRLDOS for a variety of interatomic distances. Solid lines in figure \ref{fig:On_site_fits_exponential} show first moments data for $s$, $p$ and $d$ LDOS computed for bulk FCC Cu using the QuantumWise Atomistix Tool Kit (ATK) DFT package \cite{atk, PhysRevB.65.165401, soler2002siesta} for several bulk lattice parameters. The functional of Perdew, Burke and Ernzerhof (PBE) \cite{perdew1996generalized} in the Generalized Gradient Approximation (GGA) was used for exchange and correlation in DFT. Although GGA/PBE has been used in this paper, the method presented is sufficiently general that any functional suitable for the problem at hand may be used. Once this data is generated, an expression for the variation of on-site elements with interatomic distances can then be fit to the first moments data rather than assuming an \textit{ad} \textit{hoc} expression \textit{a} \textit{priori}.

A number of functional forms can be fit to this data, but a simple exponential functional form shown below was adopted for the materials considered in this paper. This form is shown below.
\begin{equation}
\label{eq:onsites}
E_{il} = \epsilon_{il}+\sum\limits_{j=NN} k_{l}\exp[{-p_{l}(\frac{R_{ij}}{R^{(0)}}-1)}], l=s, p, d
\end{equation}
Where $\epsilon$ is a constant and $p$ is a decay exponent corresponsing to angular momentum $l$.

If intra-atomic terms are included in the Hamiltonian, equation \ref{eq:onsites} takes the form of a diagonal (i.e., on-site, same-orbital) matrix element including the effects of strain. This form suggests the following \textit{Ansatz} for the strain-dependence of the intra-atomic integrals,

\begin{equation}
\label{eq:intraatomicstrain}
I_{ll'm} = {I^{(0)}}_{ll'm}\exp[-p_{ll'm}(\frac{R_{ij}}{{{R_{ij}}^{(0)}}}-1)]
\end{equation}
where the superscript $(0)$ denotes unstrained bond length. The \textit{Ansatz} suggested above was found to be suitable for the elements considered in this paper and gives accurate fits for a wide variety of environments as shall be described in detail later in the paper. The procedure outlined above can be repeated for each individual element in the system to obtain a model for the environment-dependence of the intra-atomic elements. 

The deep insight that the Moments theorem gives is that it shows how to determine the environment dependence of the intra-atomic terms through information already obtained \textit{ab} \textit{initio}. Here the environment-variation of the intra-atomic terms is firmly rooted in physics: It follows directly from the behavior of their \textit{ab} \textit{initio} counterparts. The intra-atomic elements $I$ and their corresponding decay exponents $p$ are the only additional parameters required in writing the TB Hamiltonian.

\begin{figure}[H]
	\centering
		\includegraphics[width=7.0in]{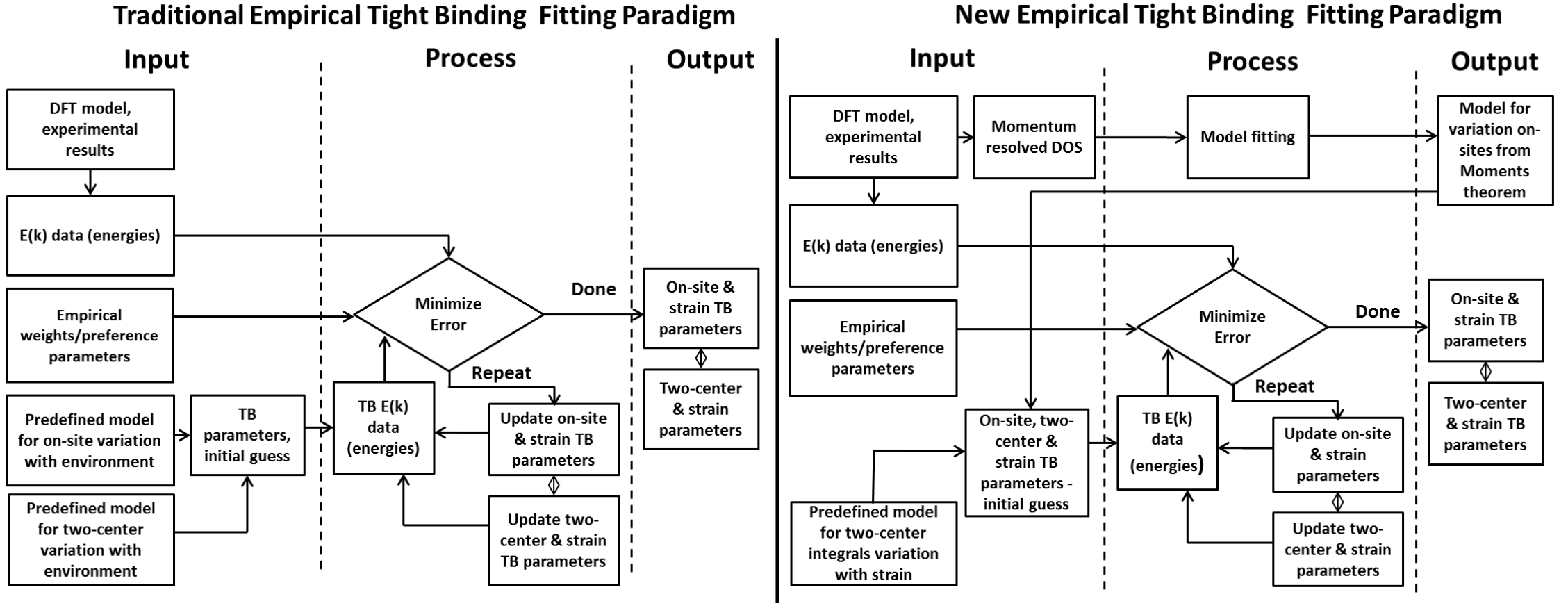}
	\caption{Process flowchart for traditional and new TB paradigm. A significant change in the new paradigm is that instead of an \textit{a} \textit{priori} assumed model for variation of on-sites with strain, a model is explicitly obtained by appealing to the Moments Theorem}	
	\label{fig:flowcharts}
\end{figure}

The new TB method described above thus represents an important change in the traditional empirical TB paradigm. As shown in figure \ref{fig:flowcharts}, the traditional TB paradigm is characterized by a pre-defined model for variation of on-sites and two-center integrals with changes in environment. The new TB fitting paradigm retains the TB model for variation of two-center elements since it is physically justified. The predefined and often arbitrary models used in the literature for on-sites, however, are replaced by a physically justified model based on the link between on-site elements and the first moment of the MRLDOS. The model for the variation of the on-site energies has therefore become the output of a specific DFT prediction, rather than an \textit{a} \textit{priori} assumption.

\section{TB parameterization - Fitting process}
\subsection{DFT target data}
Band structure targets were computed in DFT using the ATK toolkit and GGA-PBE functional. A double-zeta polarized local basis of $s$, $p$ and $d$ orbitals was used in generating DFT data. A real space grid spacing of 0.25 Bohr was used for the self-consistent solution of the Kohn-Sham and Poisson equations. Monkhorst-Pack k-space grids \cite{monkhorst1976special} of (8X8X8) k-points were found to give total energy convergence of $10^{-4}$ eV in bulk for FCC Cu, Au, Ag and Al.

Bulk, unstrained band structures for FCC Cu, Au, Al and Ag primitive unit cells at their experimental lattice parameters were then generated along a variety of paths joining high-symmetry points in the Brillouin Zone (BZ). 101 k-points per energy band and a total 9 bands were used as target bulk band structures per element.

A variety of homogeneous and inhomogeneous strained band structure targets were also generated. For homogeneous strain, band structure targets were generated in a similar fashion to bulk targets but with different lattice parameters. Up to 10\% compressive and tensile strain was used to generate homogeneous strain targets. To capture the effect of inhomogeneous strain, non-primitive unit cells along different crystal orientations were generated. These unit cells were then strained uniaxially and biaxially. Arbitrary strains were used to test the strength of the model. Up to 6\% uniaxial and bi-axial strain was used to generate strained DFT target band structures.

The same procedure outlined for bulk materials was used to generate targets for nanostructures. Nanowires having different cross sectional areas and transport orientations were created in DFT and their band structures were generated. Band structures for 10-20 nanowires per element having [001], [110] and [111] orientations and homogeneous and inhomogeneous strain were included in fitting targets. 

\subsection{TB basis and fitting algorithm}

A 9-orbital TB basis consisting of one $s$, three $p$ and five $d$ orbitals was used. For fitting the TB parameters, a constrained version of the Levenberg-Marquardt non-linear least squares algorithm \cite{more1978levenberg} was used. 
A weighted sum of residual errors between TB and DFT band structure energies was used as a measure of fitness of the parameter set. Higher weights were given to bands at or below the Fermi Level (FL).

It is important to highlight an important difference between TB parameterization for semiconductors and metals. In semiconductors, once a doping concentration is decided, the FL can be fixed in energy. Usually, effective masses at high symmetry points, energy gaps and certain key energies like the conduction band minimum (CBM), valence band maximum (VBM) are fit to high accuracy. The rest of the bands are given lower weights in the fitting process. In metals, in the absence of doping and passivation in nanostructures, the FL must be calculated by occupying bands with electrons. Thus it is important from a fitting perspective that all bands below and near the FL need to be well fit.

Two sets of parameters have been generated as part of this work. The first parameter set is a First Nearest Neighbor (1NN) Orthogonal parameter set for FCC Cu, Au, Al and Ag. This parameter set is suitable for use in situations where atomic coordination is bulk-like - bulk materials with or without strain, alloys, hetero-metal lattice matched interfaces. The other parameter set is a Second Nearest Neighbor (2NN)Orthogonal parameter set for Cu. This set fits bulk targets in addition to strained and unstrained nanostructure band structure targets where atomic coordination is far from bulk: for instance, in grain boundary structures, thin films and nanowires. A total of 43 independent TB parameters is required per metal in the new model. Tables \ref{table:parameters_1NN} and \ref{table:parameters_2NN} in the appendix A and B contain the two TB parameter sets. 

20 band structure targets containing about 20000 target energy points per metal were generated in DFT and fitted using the present model.  Alloy and hetero-metal interface band structure targets were not explicitly included in the fitting process. Simple averaging from respective bulk TB parameters was used to determine the interaction parameters between atoms of dissimilar metals in alloy and interface electronic structure calculations. This approximation was undertaken to keep the number of target band structures manageable, since the explicit inclusion of a variety of alloy and interface band structure data would significantly increase the number of targets to be fitted.

Physical quantities such as the temperature averaged DOS, position of Fermi Level, and approximate resistivity computed in the new TB model compared against DFT for bulk FCC Cu, Au, Ag and Al in table \ref {table:physquant} 

\section{Model Validation}
In this section, key results for electronic band structure and transport comparing DFT and TB obtained in the present model are shown. Due the large number of target band structures used in the fitting process, only representative results are shown in the main body of the paper in following sub-sections. Additional results are included the appendix.

\subsection{Bulk electronic structure}

\begin{figure}[H]
	\centering
		\includegraphics[width=7.0in]{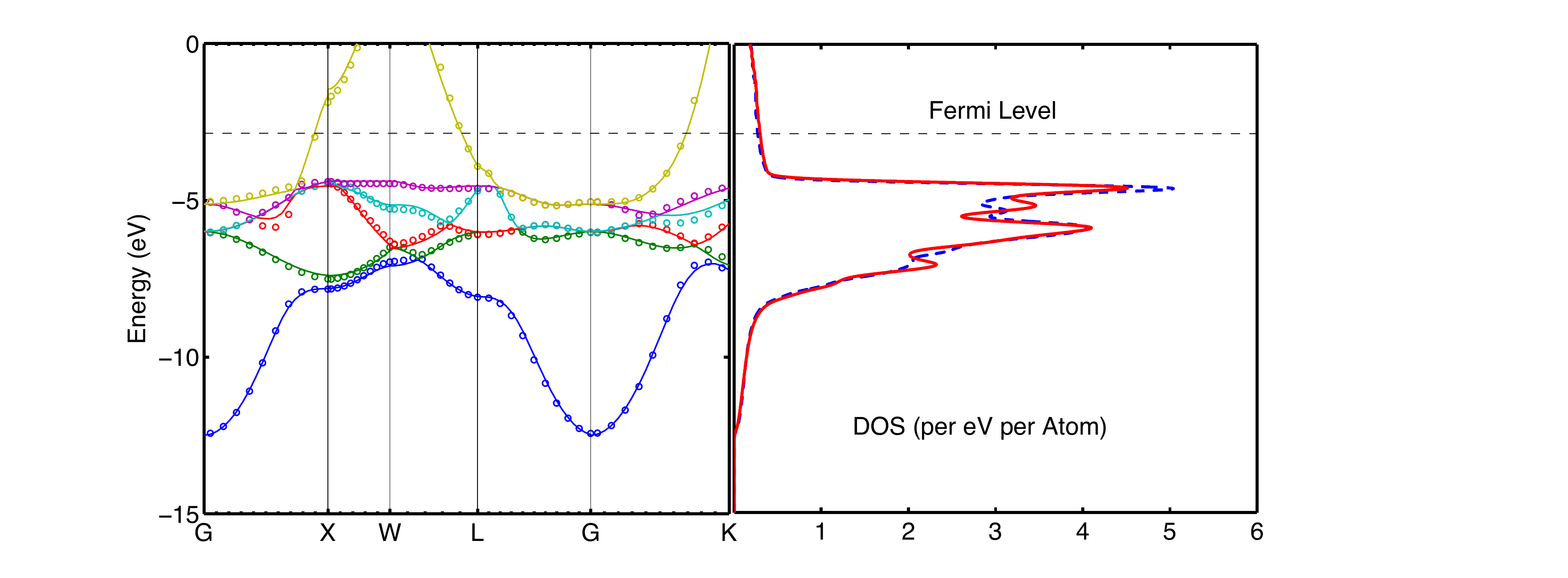}
	\caption{Bulk band structure of Cu at the experimental lattice parameter of $a_0 = 3.61 \AA$ obtained using the TB parameters optimized in the new model (circles) compared to the corresponding DFT band structure (solid lines). Also plotted on the same energy axis for comparison is the TB DOS (dashed line) and DFT DOS (solid line)}	
	\label{fig:Cu_bulk}
\end{figure}

Bulk TB band structure and DOS computed for Cu in the present model using 1NN parameters at the experimental FCC lattice parameter of $a_0 = 3.61 \AA$ and its comparison to its DFT counterpart is shown in figure \ref{fig:Cu_bulk}. Additional band structure and DOS comparisons between TB and DFT bulk band structures for Au, Al and Ag are included in the appendix.
The DOS in figure \ref{fig:Cu_bulk} was obtained by using a Gaussian broadening of 0.1eV and a Monkhorst-Pack Grid of (20X20X20) k-points for both DFT and TB.
An excellent quantitative match to DFT is obtained within the new TB model. Key transport parameters such as the position of the Fermi Level, temperature averaged DOS around the Fermi Level at 300K and the resistivity computed using the formalism outlined in Appendix C are included in table \ref{table:physquant} for a more quantitative comparison. It is evident that these key physical quantities computed in TB have excellent quantitative agreement with their DFT counterparts.
\begin{table}[H]
\caption{Comparison of physical quantities computed using DFT and the new TB model for bulk FCC metals}
\centering
\begin{tabular}{c c c c c c c}
\hline \hline
Material & \multicolumn{2}{c}{DOS ($ev^{-1}atom^{-1}$)} & \multicolumn{2}{c}{Fermi Level (eV)} & \multicolumn{2}{c}{Resistivity ($\mu\Omega-cm$)} \\
 & DFT & TB & DFT & TB & DFT & TB\\
\hline \hline
Cu & 0.290 & 0.275 & -2.817 &-2.816  & 2.23 & 2.29 \\
Ag & 0.257 & 0.243 & -2.710 & -2.655 & 2.10 & 2.10 \\
Au & 0.260 & 0.244 & -2.140 & -2.056 & 3.15 & 3.10 \\
Al & 0.405 & 0.385 & -4.455 & -4.427 & 3.25 & 3.30 \\
\hline
\end{tabular}
\label{table:physquant}
\end{table}

\subsection{Strained electronic structure}
\begin{figure}[H]
	\centering
		\includegraphics[width=3.5in]{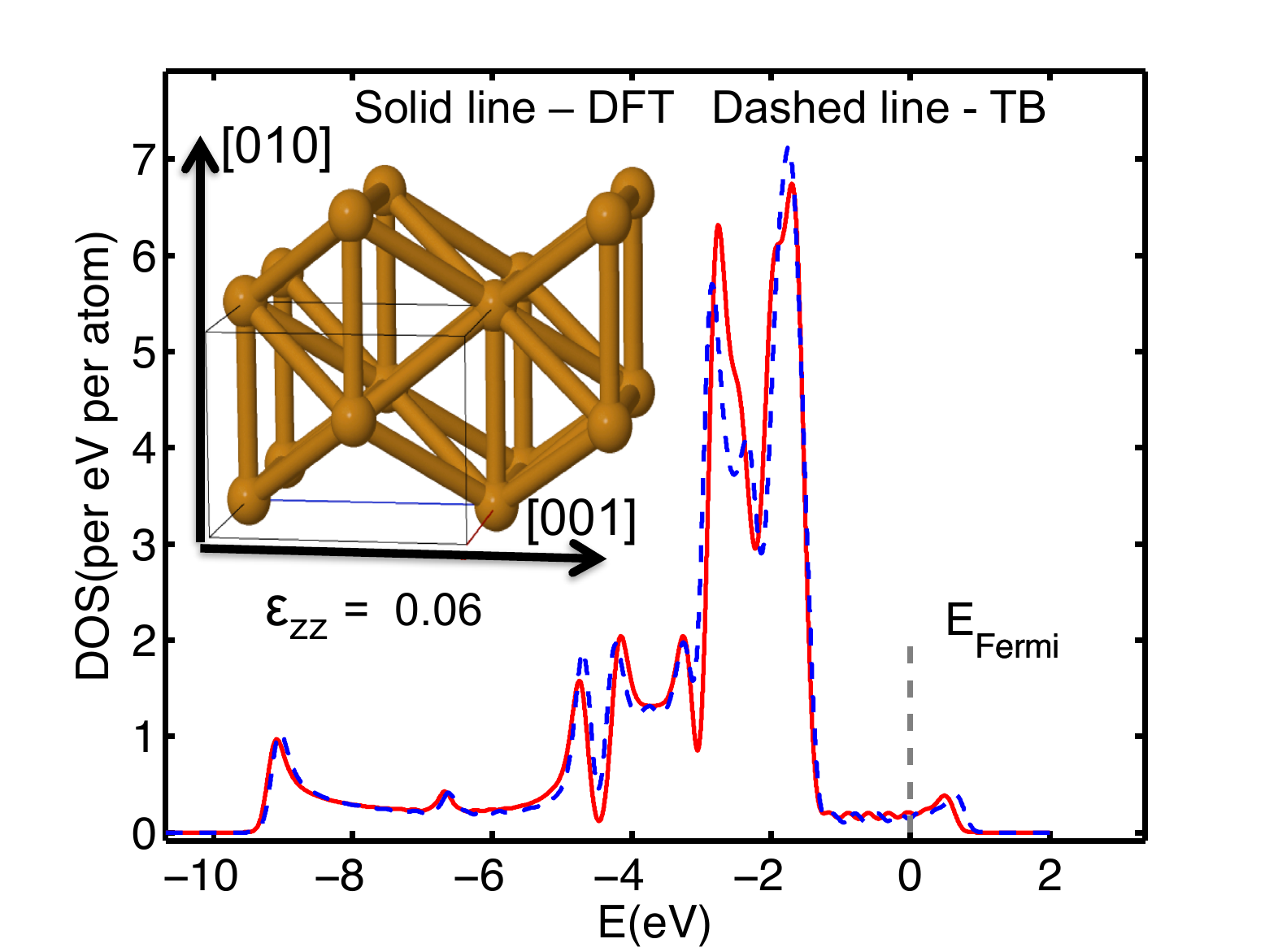}
	\caption{Unit cell structure and DOS of a Cu bulk [001] oriented unit cell strained by 6\% in the [001] direction relative to bulk lattice parameter. Plotted here are results obtained using the TB parameters optimized in the new model (dashed line) compared to the corresponding DFT band structure (solid line). The DOS corresponds to band structure from [000] to [001] normalized to $k=\pi/c$ where $c$ is the unit cell magnitude in [001]}	
	\label{fig:Cu_uniax_structure_and_bs}
\end{figure}

Figure \ref{fig:Cu_uniax_structure_and_bs} shows the unit cell structure and DOS when 6\% tensile uniaxial strain is applied along [100] crystal orientation in Cu. The unit cell was aligned with the $xyz$ coordinate system so that the transport orientation is along the $z$ axis. The DOS has been computed from the bulk band structure for a single transverse k point at $k_x=0$ and $k_y=0$. Since the coordination is still bulk-like, 1NN parameters are used to generate the comparison. Additional representative results for homogeneous and inhomogeneous strain for Au, Ag and Al are included in the appendix. It is evident that a good qualitative and quantitative match between the TB and DFT DOS data is obtained for strained targets.
The environment dependence and implicit self-consistency in the present model ensure that same 1NN parameters fit homogeneous and inhomogeneous strained targets accurately.

\subsection{Alloy electronic structure}
Band structures for solid-substitutional alloy supercells across a wide range of compositions were generated and compared with DFT for the Au-Ag system . A lattice parameter corresponding to the average of the individual metals was used to generate DFT data for all systems. Since the mismatch between the lattice parameters is less than 0.1 \%, this is not an unreasonable assumption.

Au-Au and Ag-Ag interactions remain unchanged from bulk. Since coordination is still bulk-like, 1NN parameters were used in generating the comparison. Au-Ag interactions were averaged from Au-Au and Ag-Ag bulk values, as mentioned previously in the section on the details of the fitting process. It is evident that even by averaging interaction TB parameters for this material combination, a good qualitative and quantitative match is obtained with DFT. This attests to the quality of the repsective bulk parameterizations.

\begin{figure}[H]
	\centering
		\includegraphics[width=3.5in]{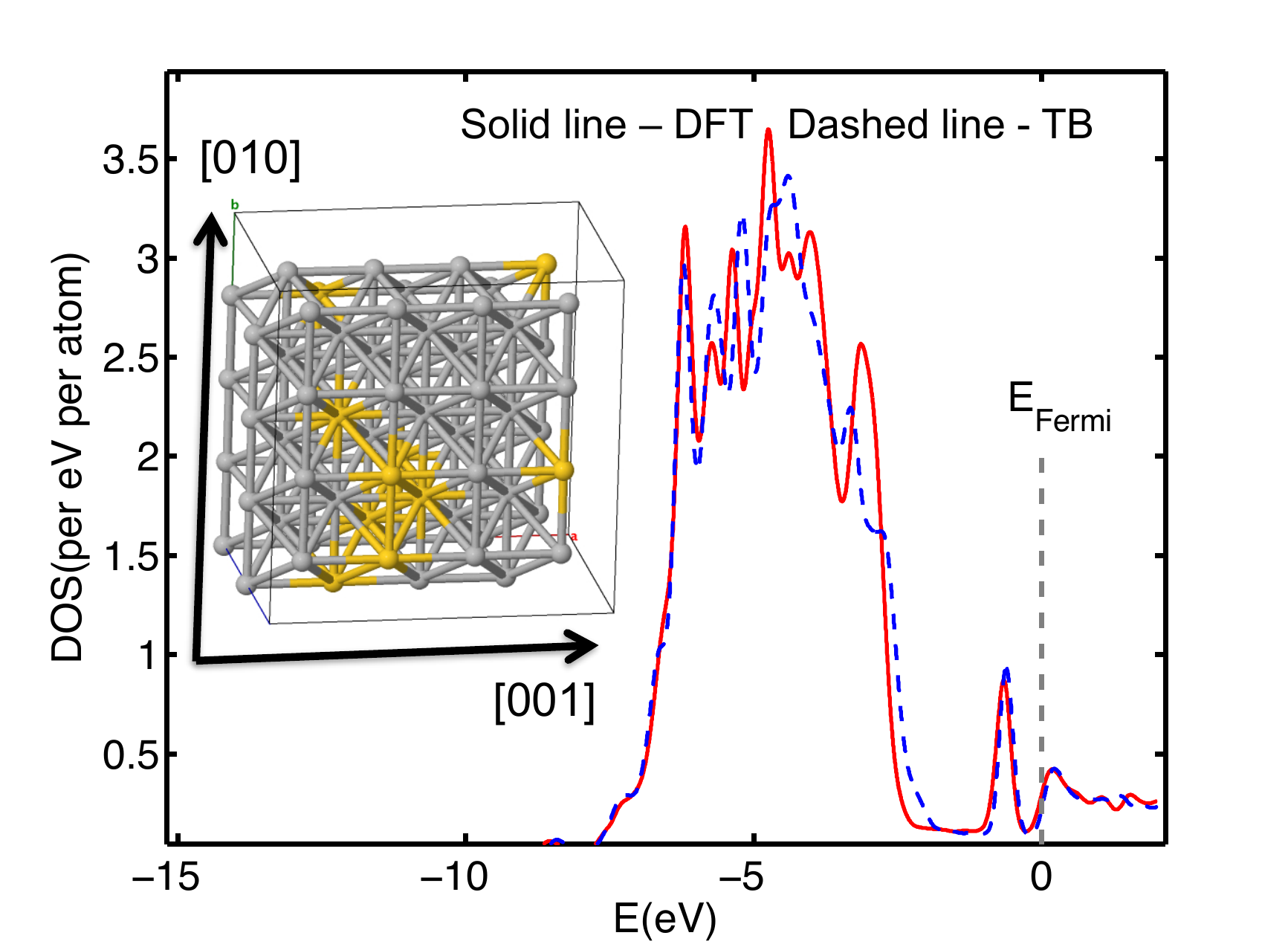}
	\caption{ Atomic Structure and DOS of ordered 19\% Ag and 81\% Au alloy supercell obtained using TB parameters optimized in the new model (dashed lines) compared to the corresponding DFT DOS (solid lines). The DOS data was obtained from band structure for a single transverse k point. Silver-colored spheres represent Ag atoms while the golden spheres represent Au atoms.}	
	\label{fig:Au19Ag81_structure_and_bs}
\end{figure}

Figure \ref{fig:Au19Ag81_structure_and_bs} shows the atomic structure and DOS for 19\% Au and 81\% Ag alloy supercell containing 64 atoms and the corresponding DFT target DOS. Similar to the case of bulk strain, the DOS has been generated from supercell band structure data for a single transverse k-point. Additional alloy DOS results for the Au-Ag system are included in the appendix.
Similar to the case of bulk strain for the elementary metals, environment dependence and implicit self-consistency in present TB model ensure that same 1NN parameters fit alloy supercell band structures accurately. It is anticipated that explicit inclusion of alloy targets in the fitting process may improve the quantitative match even further than that obtained using simple averaging of interactions between dissimilar atoms.

\subsection{Metal-metal interface electronic structure}
\begin{figure}[H]
	\centering
		\includegraphics[width=3.5in]{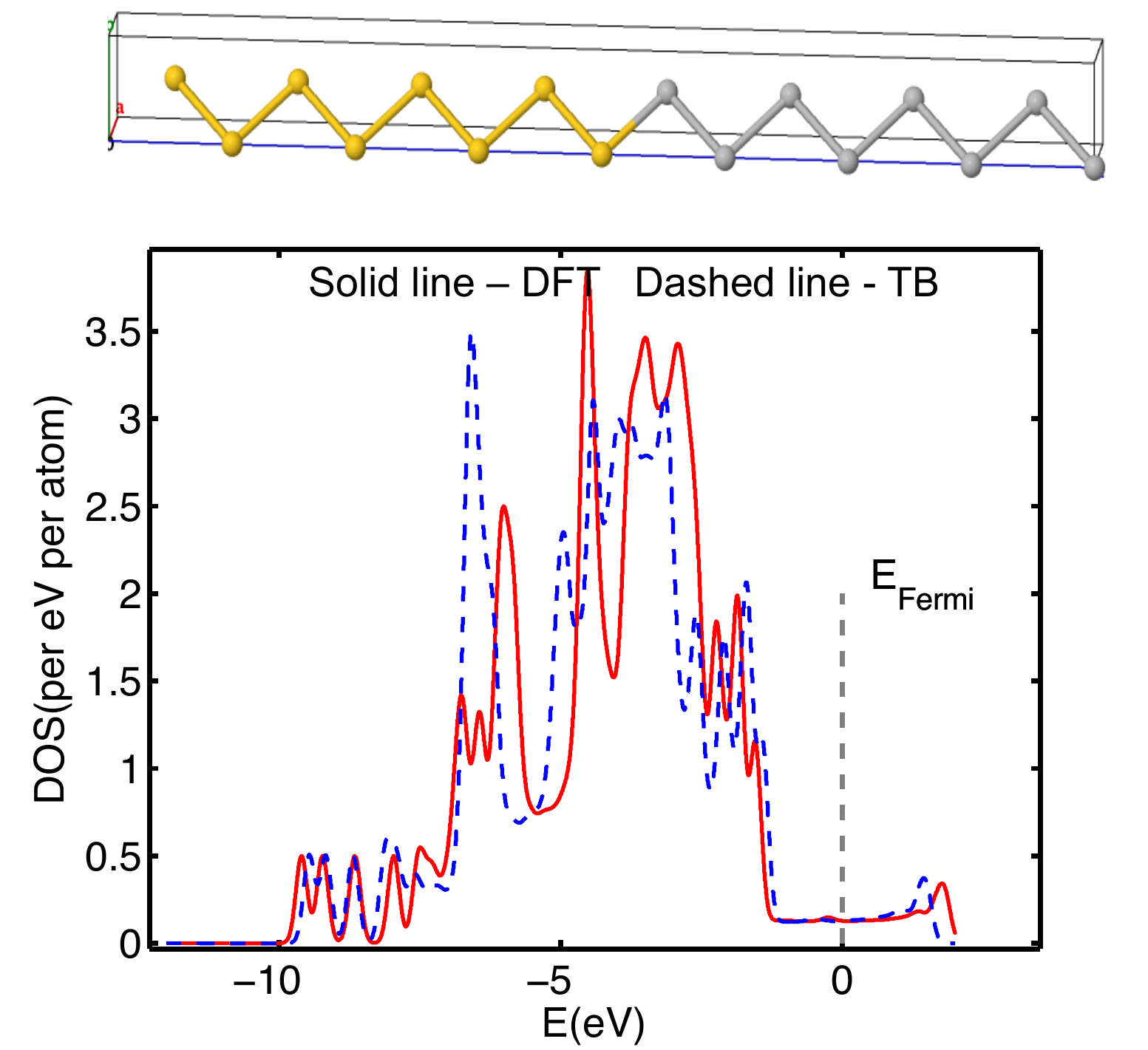}
	\caption{ Atomic Structure and DOS of Ag-Au superlattice unit cell obtained using TB parameters optimized in the new model (dashed lines) compared to the corresponding DFT DOS (solid lines). The DOS data was obtained from band structure for a single transverse k point. Silver-colored spheres represent Ag atoms while the golden spheres represent Au atoms.}	
	\label{fig:AuAg8Layers_interim}
\end{figure}

The procedure outlined above for the Ag-Au alloy structures were also followed for a number of Au-Ag superlattice band structures. Au-Ag interactions were averaged from their respective bulk values and the superlattice band structure was computed. This band structure information is converted to DOS information and compared to DFT results. Figure \ref{fig:AuAg8Layers_interim} shows one such result for DOS obtained from band structure corresponding to a single transverse k-point of an Au-Ag superlattice unit cell 16 atomic layers thick in the direction of growth.
The overall TB DOS profile shows reasonable qualitative agreement with DFT, but is mismatched at certain energies. Around the Fermi Level, still, the DOS calculated in TB is mismatched from DFT by less than 2\% indicating excellent quantitative agreement. Even in the absence of explicit fitting of interface band structure targets, it can be seen that the same bulk 1NN parameters give an electronic structure that shows a quantitative agreement with the self-consistent electronic structure computed in DFT around energies of interest. 

\subsection{Nanostructures}
\begin{figure}[H]
	\centering
		\includegraphics[width=3.5in]{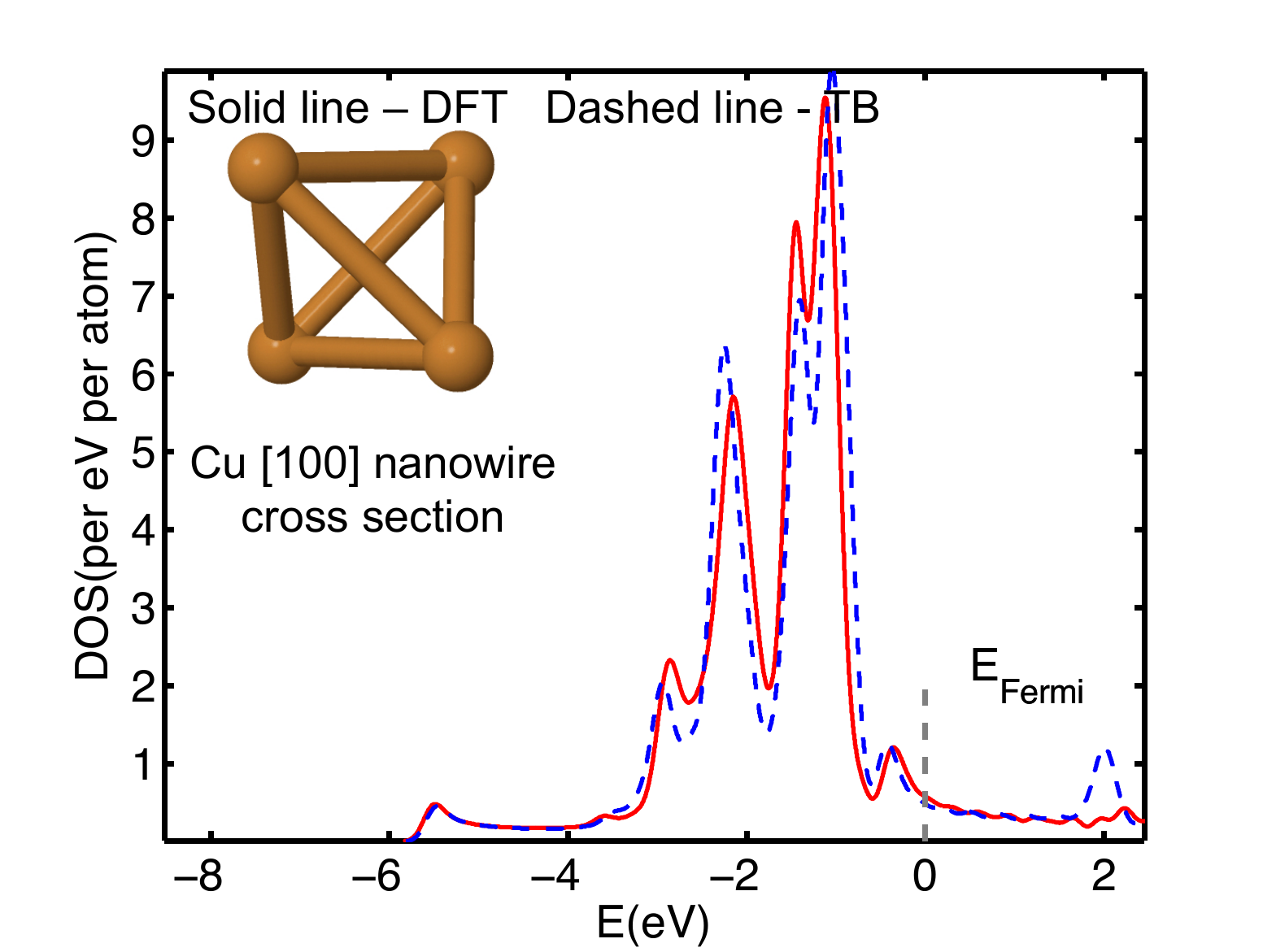}
	\caption{ DOS of Cu [100] oriented cross section with just 4 atoms in the unit cell pictured, obtained using TB parameters optimized in the new model (Dashed lines) compared to the corresponding DFT band structure (solid line). The [100] direction is perpendicular to the plane of the paper and the unit cell is periodic in this direction}	
	\label{fig:Cu_100_bs_and_structure}
\end{figure}

Figure \ref{fig:Cu_100_bs_and_structure} shows the atomic structure and DOS comparison for a Cu [100] oriented nanowire. 2NN parameters in table \ref{table:parameters_2NN} were used to generate the TB DOS. The structure has just 4 Cu atoms in the cross section. Consequently, the atoms in the structure see a very different bonding environment than bulk atoms. Additional results for Cu nanowires of different orientations and different bonding environments, but generated using the same 2NN parameters are included in the appendix.  
Since the same 2NN parameters fit bulk and nanowire band structures well, it is evident that the new TB model results in improved transferability due to its environment-aware nature and implict self-consistency.

\subsection{Electronic transport using the new model}
\begin{figure}[H]
	\centering
		\includegraphics[width=3.5in]{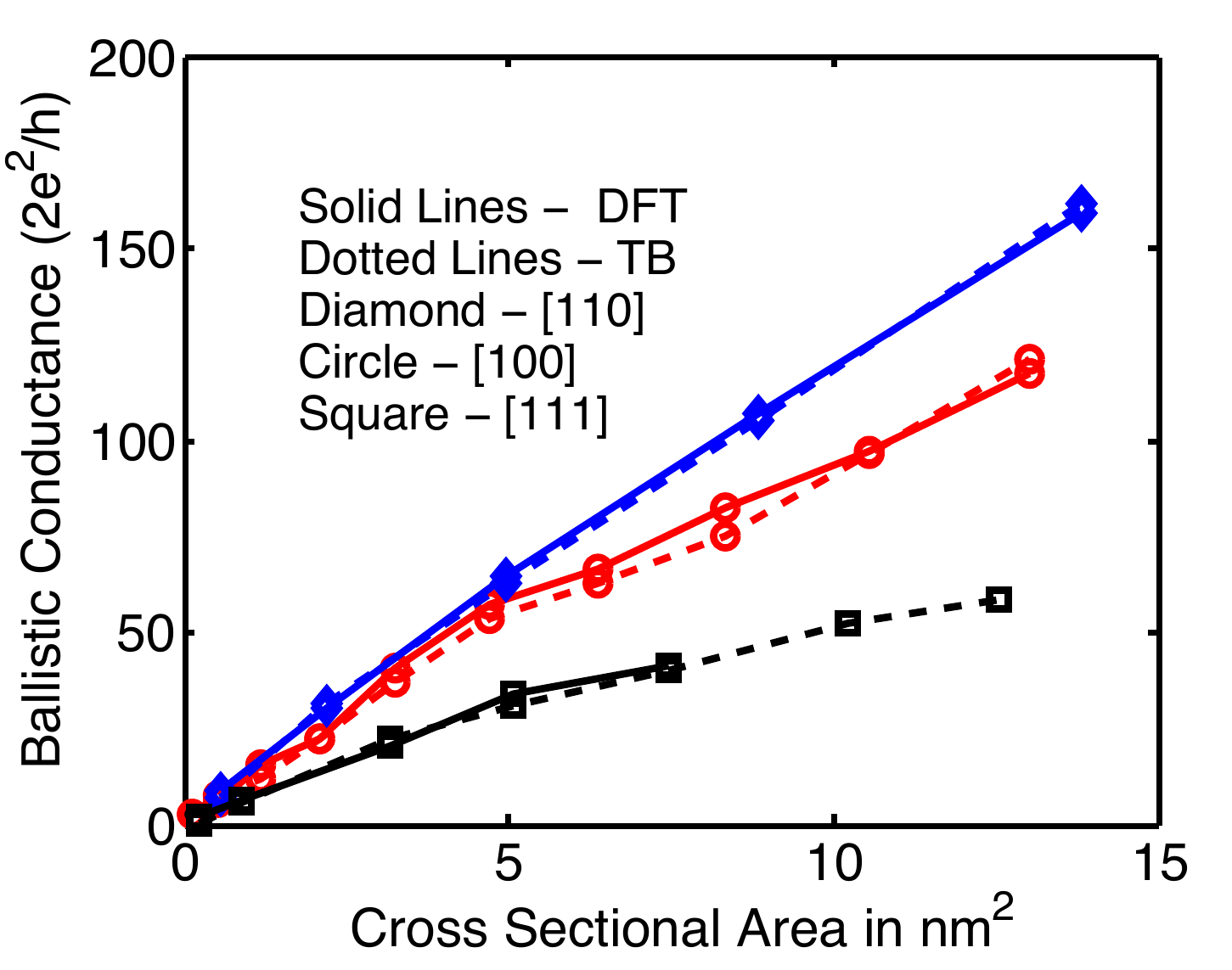}
	\caption{Ballistic conductance of square Cu nanowires of different orientations computed using the TB model and compared to ballistic conductance computed using DFT}	
	\label{fig:wiretransmissionversusarea}
\end{figure}

The new TB model and parameters have been incorporated into the Nanoelectronic Modeling Suite (NEMO5) electronic transport simulator \cite{steiger2011nanoelectronic, fonseca2013efficient}. In order to compare the accuracy of the new TB parameterization with DFT for electronic transport, ballistic conductance for square Cu nanowires of different transport orientations was computed using the 2NN parameters in table \ref{table:parameters_2NN} and compared with DFT results obtained using the ATK package. The procedure was repeated for a variety of cross sectional areas. Figure \ref{fig:wiretransmissionversusarea} indicates that the ballistic conductances computed in the new TB model show an excellent qualitative and quantitative match with their DFT counterparts.


\section{Conclusion}

In conclusion, a new TB model suitable for electronic transport in metals has been presented in this paper. Environment-awareness is incorporated in the model through the use of intra-atomic diagonal and off-diagonal elements. In contrast with existing TB models of metals, this model uses the relationship between intra-atomic matrix elements and first moments of MRLDOS to incorporate environment-dependence in a physically transparent way. 

The TB model has been validated against DFT calculations of electronic structure and transport and is transferable to bulk FCC metals with and without strain, binary metal alloys, hetero-metal interfaces and metallic nanostructures. To our knowledge, this is also the first instance where bulk metallic electronic structure is reproduced accurately within an orthogonal, first nearest neighbor TB model. In a subsequent paper the model is applied to a systematic analysis of the effect of confinement and homogeneous strain on the conductance of Cu.

The model and its exploration are fully implemented in the Nanoelectronic Modeling Suite NEMO5 \cite{fonseca2013efficient, steiger2011nanoelectronic}, which is released under an academic open source license on nanoHUB \cite{NEMO5support}

\appendix

\section{1NN parameters for Cu, Al, Au and Ag}
The table below lists the First nearest neighbor Two-Center Integrals and Intra-Atomic Integrals in the Slater-Koster format. The Two-Center Integrals are denoted by the letter $V$ while the Intra-Atomic Integrals are denoted by $I$. The strain dependence of the Two-Center Integrals is captured by the exponents $q$ while the strain dependence of the Intra-Atomic Integrals is captured by the exponent $p$.  $R_{0_{inter}}$ is the reference zero strain length for Two-Center Integrals while $R_{0_{intra}}$ is the reference zero-strain length for Intra-Atomic integrals. The Integrals are in units of eV, the exponents unitless and the distances are in units of Angstrom.

	
\begin{longtable}{c c c c c}
	\hline\hline
	Parameter& Cu & Ag & Au & Al \\
	\hline \hline

	$\epsilon_s$ & -4.3325   &  -3.5253& -7.0609& -6.0263\\
	$\epsilon_p$ & 0.8458    & 1.4098& 0.9004& -6.7152\\
	$\epsilon_d$ & -3.3164   & -5.5780& -5.4325& 6.4493\\ \\

	$V_{ss\sigma}$ & -0.9992 & -0.8864& -0.9261& -0.7075\\
	$V_{sp\sigma}$ & 1.4060  & 1.2238& 1.3669& 1.1799\\
	$V_{sd\sigma}$ & -0.5171 & -0.5268& -0.6941& -0.9986\\
	$V_{pp\sigma}$ & 1.9544  &  1.5428& 1.7926& 2.2634\\
	$V_{pp\pi}$    & -0.5788 & -0.5098& -0.5155& -0.1741\\
	$V_{pd\sigma}$ & -0.5264 & -0.6058& -0.9479& -2.3998\\
	$V_{pd\pi}$    & 0.2537  & 0.1868& 0.2972& 0.3417\\
	$V_{dd\sigma}$ & -0.3642 & -0.4540& -0.6844& -3.1204\\
	$V_{dd\pi}$    & 0.2464  & 0.2456& 0.3381& 1.0852\\
	$V_{dd\delta}$ & -0.0598 & -0.0496& -0.0592& -0.0949\\ \\

	$I_{ss\sigma}$ & 0.3222 & 0.2888& 0.5000& -0.0902\\
	$I_{sp\sigma}$ & -13.4601  & 27.5677& 47.4232& 47.3913\\
	$I_{sd\sigma}$ & 0.0012 & -0.5037& 0.0339& -1.5323\\
	$I_{pp\sigma}$ & 1.5605  & 0.3040& 1.0660& -0.0016\\
	$I_{pp\pi}$    & -0.1228 & 0.2953& 0.1271& 1.0054\\
	$I_{pd\sigma}$ & -10.2694 & 99.9958& 99.8976& -8.9001\\
	$I_{pd\pi}$    & -3.8434  & -99.1472& -98.6434& 9.6930\\
	$I_{dd\sigma}$ & -0.2348 & 0.7252& 0.9232& 3.9694\\
	$I_{dd\pi}$    & -0.2498  & 0.2511& 0.3693& 1.4278\\
	$I_{dd\delta}$ & -0.1199 & -0.9280& -0.9722& -2.0288\\ \\
 
	$q_{ss\sigma}$ & 2.3603 & 2.5004& 3.4147& 1.7924\\
	$q_{sp\sigma}$ & 2.0827 & 1.7035& 3.0152& 1.9547\\
	$q_{sd\sigma}$ & 3.0696 & 3.8742& 4.0517& 1.5753\\
	$q_{pp\sigma}$ & 2.4409 & 1.4366& 2.5772& 2.1836\\
	$q_{pp\pi}$    & 2.6721 & 4.0872& 2.9059& 0.2495\\
	$q_{pd\sigma}$ & 4.3038 & 5.3603& 4.2849& 1.9492\\
	$q_{pd\pi}$    & 5.1106 & 6.7768& 4.6552& 0.0000\\
	$q_{dd\sigma}$ & 4.8355 & 5.5990& 5.4403& 1.4962\\
	$q_{dd\pi}$    & 4.7528 & 5.2114& 5.0338& 1.7435\\
	$q_{dd\delta}$ & 4.2950 & 3.9194& 2.4849& 0.1374\\ \\

	$p_{ss\sigma}$ & 2.6580  & 3.3967  & 3.4130& 9.5144\\
	$p_{sp\sigma}$ & 75.2459 & 99.1027 & 75.4943& 41.6508\\
	$p_{sd\sigma}$ & 31.1555 & 0.4269  & 1.8747& 11.19627\\
	$p_{pp\sigma}$ & 3.6612  & 3.1565  & 4.5685& 16.7744\\
	$p_{pp\pi}$    & 0.2321  & 5.1796  & 6.1124& 0.7603\\
	$p_{pd\sigma}$ & 28.0089 & 99.8750 & 99.9935& 52.9425\\
	$p_{pd\pi}$    & 4.1244  & 99.1178 & 91.1682& 29.8172\\
	$p_{dd\sigma}$ & 0.2667  & 2.3977  & 1.8897& 5.5679\\
	$p_{dd\pi}$    & 1.4597  & 2.1463  & 2.0762& 11.0296\\
	$p_{dd\delta}$ & 0.0189  & 1.7029  & 1.3285& 10.4785\\ \\

	$R_{0_{inter}}$  & 2.5526 & 2.8890& 2.8837& 2.8634\\
	$R_{0_{intra}}$  & 2.5526 & 2.8890& 2.8837& 2.8634\\

	\hline \hline


\label{table:parameters_1NN}
\end{longtable}

\subsection{Notes on the range of applicability for the parameters in table \ref{table:parameters_1NN}}
The parameters listed in table \ref{table:parameters_1NN} have been optimized using a bonding radius of $\frac{a_0}{\sqrt{2}}$. $a_0$ is varied uniformly for the case of hydrostatic strain, while it is varied non-uniformly in different crystal directions for uniaxial/biaxial strain. The maximum strain to which the parameters have been optimized to is 6\% biaxial and 10\% hydrostatic strain data in DFT. While the model developed is robust, usage of parameters beyond the data to which it has been optimized may give unphysical results, as in the case of any empirically fitted model. The parameters listed above are unsuitable for use in nanostructures where surface effects dominate and should be used for bulk-like bonding situations.

\newpage

\section{2NN parameters for Cu}
The table below lists Second nearest neighbor Two-Center Integrals and Intra-Atomic Integrals in the Slater-Koster format. The Two-Center Integrals are denoted by the letter $V$ while the Intra-Atomic Integrals are denoted by $I$. The strain dependence of the Two-Center Integrals is captured by the exponents $q$ while the strain dependence of the Intra-Atomic Integrals is captured by the exponent $p$. $R_{0_{inter}}$ is the reference zero-strain length for Two-Center Integrals while $R_{0_{intra}}$ is the reference zero-strain length for Intra-Atomic  integrals. The Integrals are in units of eV, the exponents unitless and the distances are in units of Angstrom.
\begin{longtable}{c c}

	\hline \hline
	Parameter& Value\\
	\hline \hline
	$\epsilon_s$ & -4.5236 \\
	$\epsilon_p$ & 0.1689 \\
	$\epsilon_d$ & -4.7007\\ \\

	$V_{ss\sigma}$ & -0.9588 \\
	$V_{sp\sigma}$ & 1.4063  \\
	$V_{sd\sigma}$ & -0.1841  \\
	$V_{pp\sigma}$ & 1.4025  \\
	$V_{pp\pi}$    & -0.5730 \\
	$V_{pd\sigma}$ & -0.4607 \\
	$V_{pd\pi}$    & 0.3373  \\
	$V_{dd\sigma}$ & -0.3709 \\
	$V_{dd\pi}$    & 0.2760  \\
	$V_{dd\delta}$ & -0.0735 \\ \\

	$I_{ss\sigma}$ & 0.4329 \\
	$I_{sp\sigma}$ & 0.4457  \\
	$I_{sd\sigma}$ & -0.2323 \\
	$I_{pp\sigma}$ & 1.0657 \\
	$I_{pp\pi}$    & -0.1348 \\
	$I_{pd\sigma}$ & -0.2532 \\
	$I_{pd\pi}$    & 0.0135  \\
	$I_{dd\sigma}$ & -0.2905 \\
	$I_{dd\pi}$    & 0.0002  \\
	$I_{dd\delta}$ & -0.0883 \\ \\

	$q_{ss\sigma}$ & 1.9386 \\
	$q_{sp\sigma}$ & 2.6554 \\
	$q_{sd\sigma}$ & 1.2649 \\
	$q_{pp\sigma}$ & 1.5905 \\
	$q_{pp\pi}$    & 2.9059 \\
	$q_{pd\sigma}$ & 3.8124 \\
	$q_{pd\pi}$    & 3.9330 \\
	$q_{dd\sigma}$ & 2.5348 \\
	$q_{dd\pi}$    & 3.2170 \\
	$q_{dd\delta}$ & 1.4936 \\ \\

	$p_{ss\sigma}$ & 1.9236 \\
	$p_{sp\sigma}$ & 2.8794 \\
	$p_{sd\sigma}$ & 2.8819 \\
	$p_{pp\sigma}$ & 5.5023  \\
	$p_{pp\pi}$    & 10.0272  \\
	$p_{pd\sigma}$ & 33.9945 \\
	$p_{pd\pi}$    & 82.1628  \\
	$p_{dd\sigma}$ & 10.0271  \\
	$p_{dd\pi}$    & 99.9990  \\
	$p_{dd\delta}$ & 4.8475  \\ \\

	$R_{0_{inter}}$  & 2.5526 \\
	$R_{0_{intra}}$  & 2.5526 \\

	\hline \hline


	\label{table:parameters_2NN}

\end{longtable}

\subsection{Notes on the range of applicability for the parameters in table \ref{table:parameters_2NN}}
The parameters listed in table \ref{table:parameters_1NN} have been optimized using a bonding radius of ${a_0}$. $a_0$ is varied uniformly for the case of hydrostatic strain, while it is varied non-uniformly in different crystal directions for uniaxial/biaxial strain. The maximum strain to which the parameters have been optimized to is 6\% biaxial and 10\% hydrostatic strain data in DFT. While the model developed is robust, usage of parameters beyond the data to which it has been optimized may give unphysical results, as in the case of any empirically fitted model. The parameters listed above are more suited for nanostructures, since a higher weight was given during the fitting process for low-coordination structures.

\section{Modified Landauer Scheme for resistivity computation }
First, the number of conducting modes around Fermi Level is computed. This is nothing but the ballistic conductance for periodic structures \cite{datta1997electronic}.
\begin{equation}
\label{eq:ballisticconductance}
G = \frac{2e^2}{h}\sum\limits_{n}\int{T_{n}(E)\frac{-\partial f}{\partial E} dE}
\end{equation}
where the sum is over all bands $n$ , $T(E)$ is the number of conducting modes at energy level E. The quantity $\frac{2e^2}{h}$ is the fundamental quantum of conductance. In bulk, this number needs to be computed over all transverse modes and then averaged.

From this, an approximate resistivity is computed using a modified Landauer approach \cite{zhou2008resistance}. If a mean free path $\lambda$ between scattering events exists, then the ballistic conductance $G$ computed in equation \ref{eq:ballisticconductance} above can be modified using the gollowing expression \cite{lundstrom2009fundamentals}
\begin{equation}
G' = G\frac{\lambda}{\lambda+L} \Rightarrow R' = \frac{1}{G}\frac{\lambda+L}{\lambda}
\end{equation}  
Where $L$ is the length of the conductor. 

A linear fit for the ballistic conductance versus cross sectional area (A) curve can be obtained as $G = kA+c \approx kA$ for large cross sectional areas. If the length of the conductor is much greater than the mean free path as in the case of nanowires, we have $L>>\lambda$ and we can write.
\begin{equation}
R' = \frac{1}{kA}\frac{L}{\lambda}
\end{equation}
Equating this to Ohm's Law $R = \rho\frac{L}{A}$ we get an approximate resistivity
\begin{equation}
\rho = \frac{1}{k\lambda}
\label{eq:rho_versus_slope}
\end{equation}
Thus, the resistivity computed in this approximate approach is inversely proportional to the slope of the conductance versus cross sectional area relation.


\section{Bulk band structures of Ag, Au and Al}

\begin{figure}[H]
	\centering
		\includegraphics[width=7.0in]{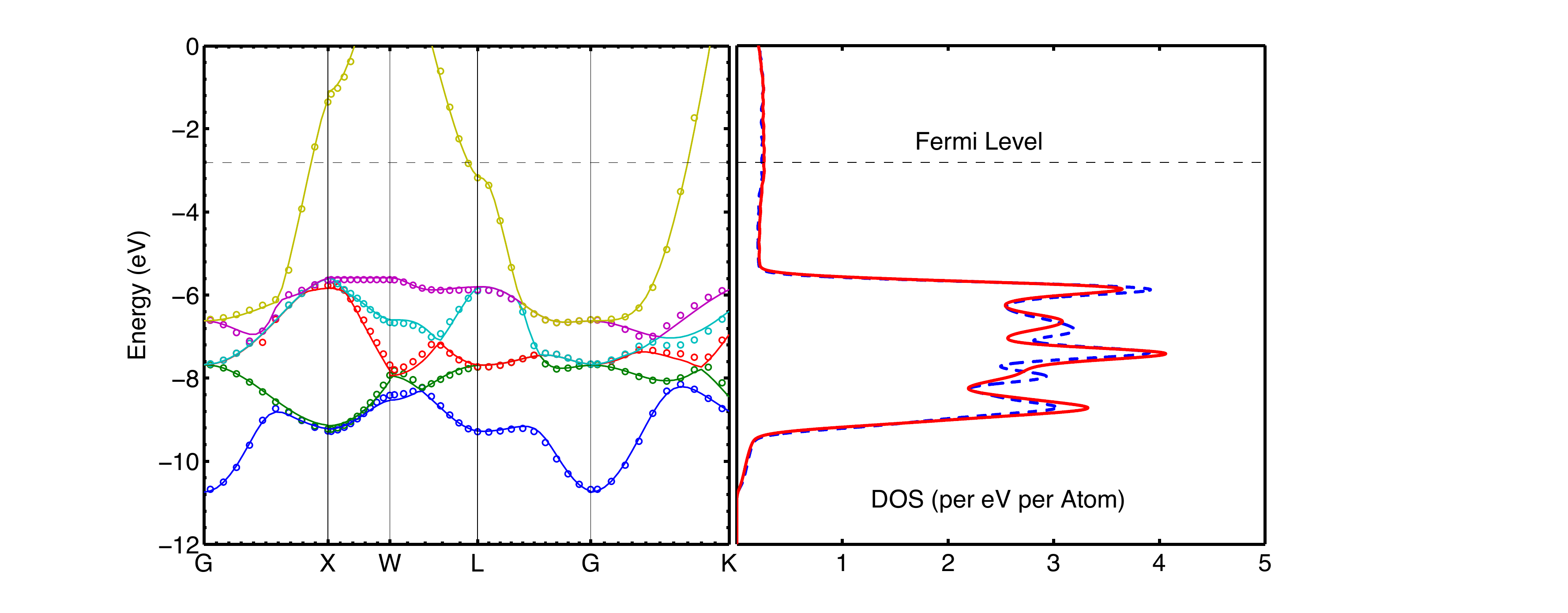}
	\caption{Bulk band structure of Ag at the experimental lattice parameter of 4.08A obtained using the TB parameters optimized in the new model (circles) compared to the corresponding DFT band structure (solid lines). Also plotted on the same energy axis for comparison is the TB DOS (dashed line) and DFT DOS (solid line)}	
	\label{fig:Ag_bulk}
\end{figure}

\begin{figure}[H]
	\centering
		\includegraphics[width=7.0in]{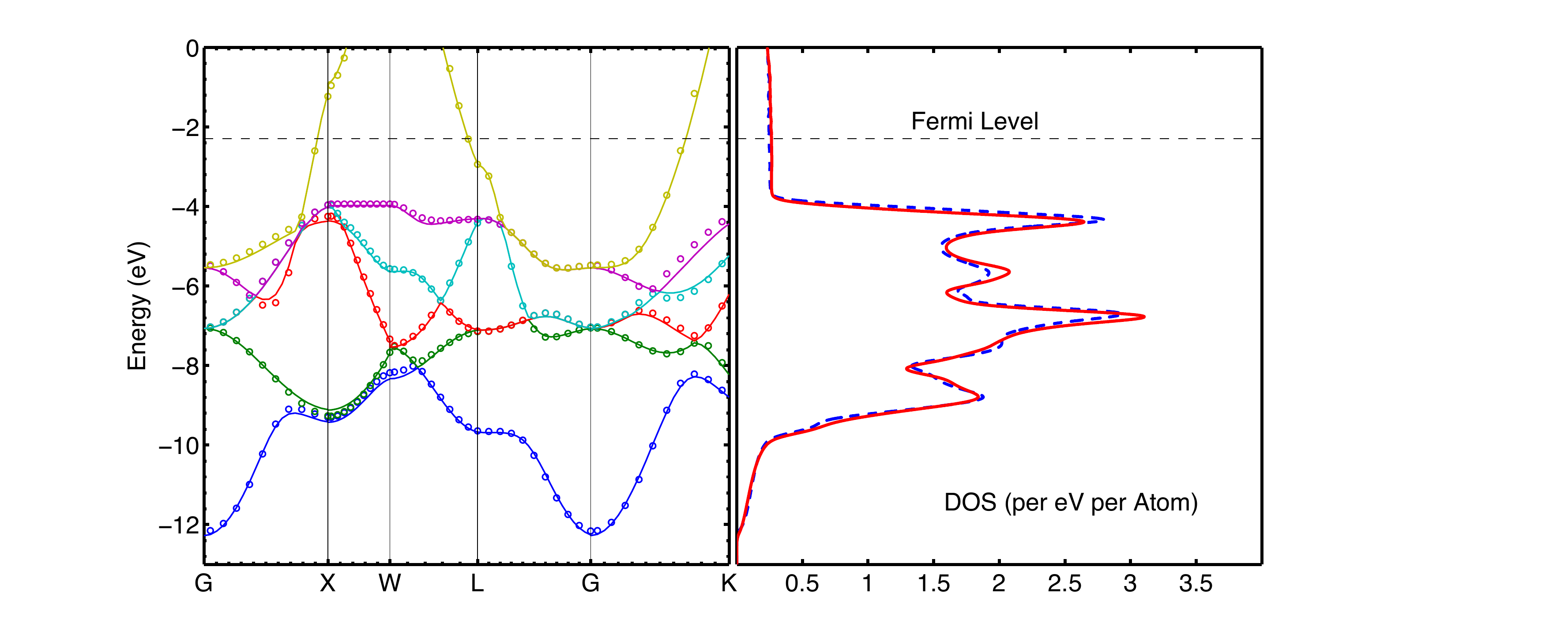}
	\caption{Bulk band structure of Au at the experimental lattice parameter of 4.078A obtained using the TB parameters optimized in the new model (circles) compared to the corresponding DFT band structure (solid lines). Also plotted on the same energy axis for comparison is the TB DOS (dashed line) and DFT DOS (solid line)}	
	\label{fig:Au_bulk}
\end{figure}

\begin{figure}[H]
	\centering
		\includegraphics[width=7.0in]{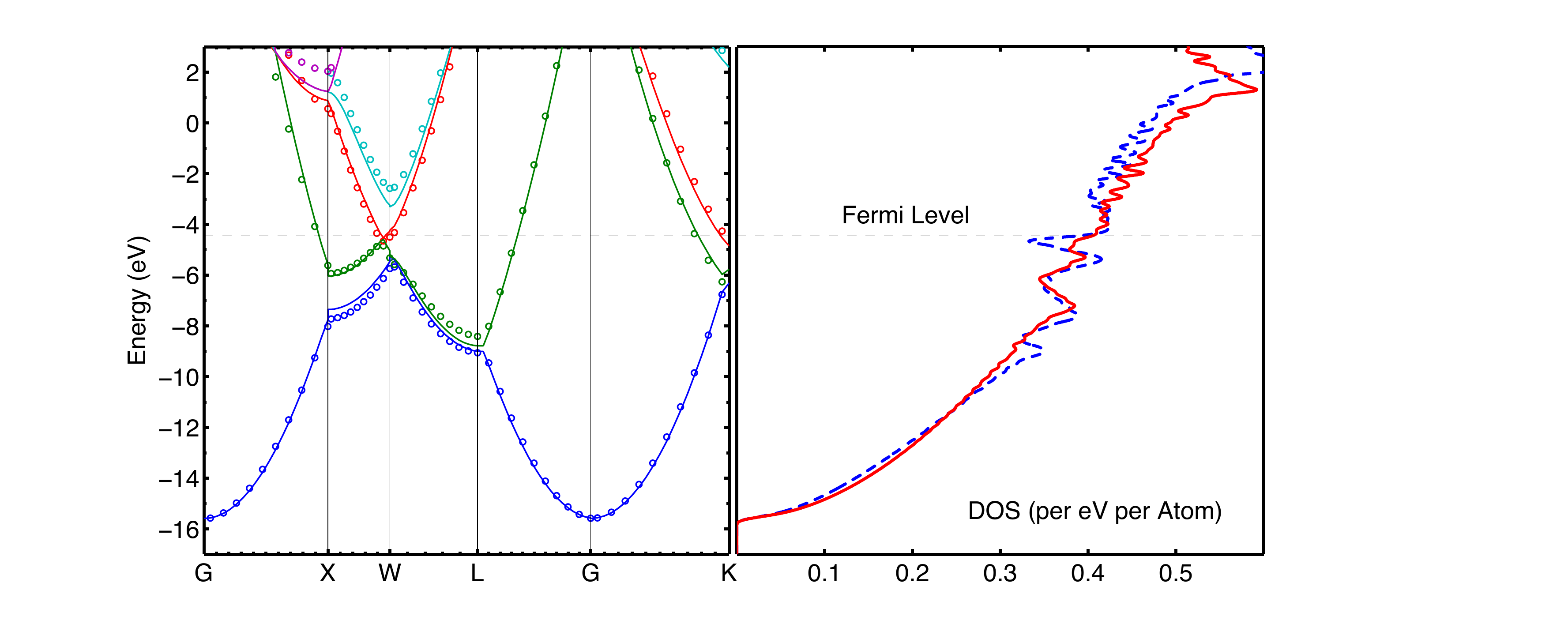}
	\caption{Bulk band structure of Al at the experimental lattice parameter of 4.05A obtained using the TB parameters optimized in the new model (circles) compared to the corresponding DFT band structure (solid lines). Also plotted on the same energy axis for comparison is the TB DOS (dashed line) and DFT DOS (solid line)}	
	\label{fig:Al_bulk}
\end{figure}

\section{Strained electronic structure of Al, Ag and Au - Examples}

\begin{figure}[H]
	\centering
		\includegraphics[width=3.5in]{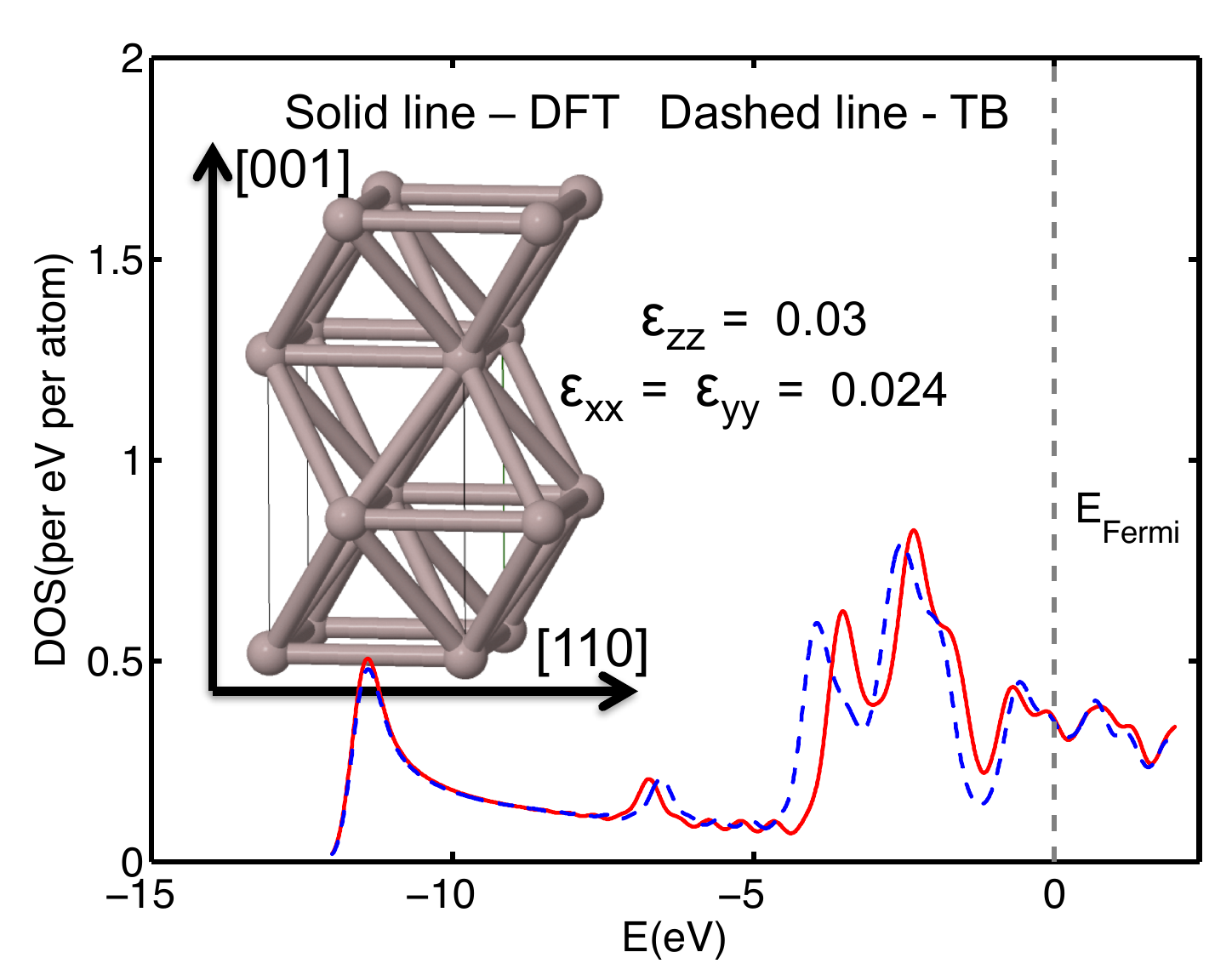}
	\caption{Unit cell structure and DOS of a Al bulk [110] oriented unit cell strained by 3 \% in the [110] direction and by 2.4 \% in the transverse plane relative to bulk lattice parameter. Plotted here are results obtained using the TB parameters optimized in the new model (dashed line) compared to the corresponding DFT results (solid line). The DOS is computed from the band structure for a single transverse k-point i.e. from [000] to [001] where the the structure is rotated so that the z-axis coincides with the [110] direction. The band structure is normalized to $k=\pi/c$ where $c$ is the unit cell magnitude in [110]}	
	\label{Al_biax_structure_and}
\end{figure}

\begin{figure}[H]
	\centering
		\includegraphics[width=3.5in]{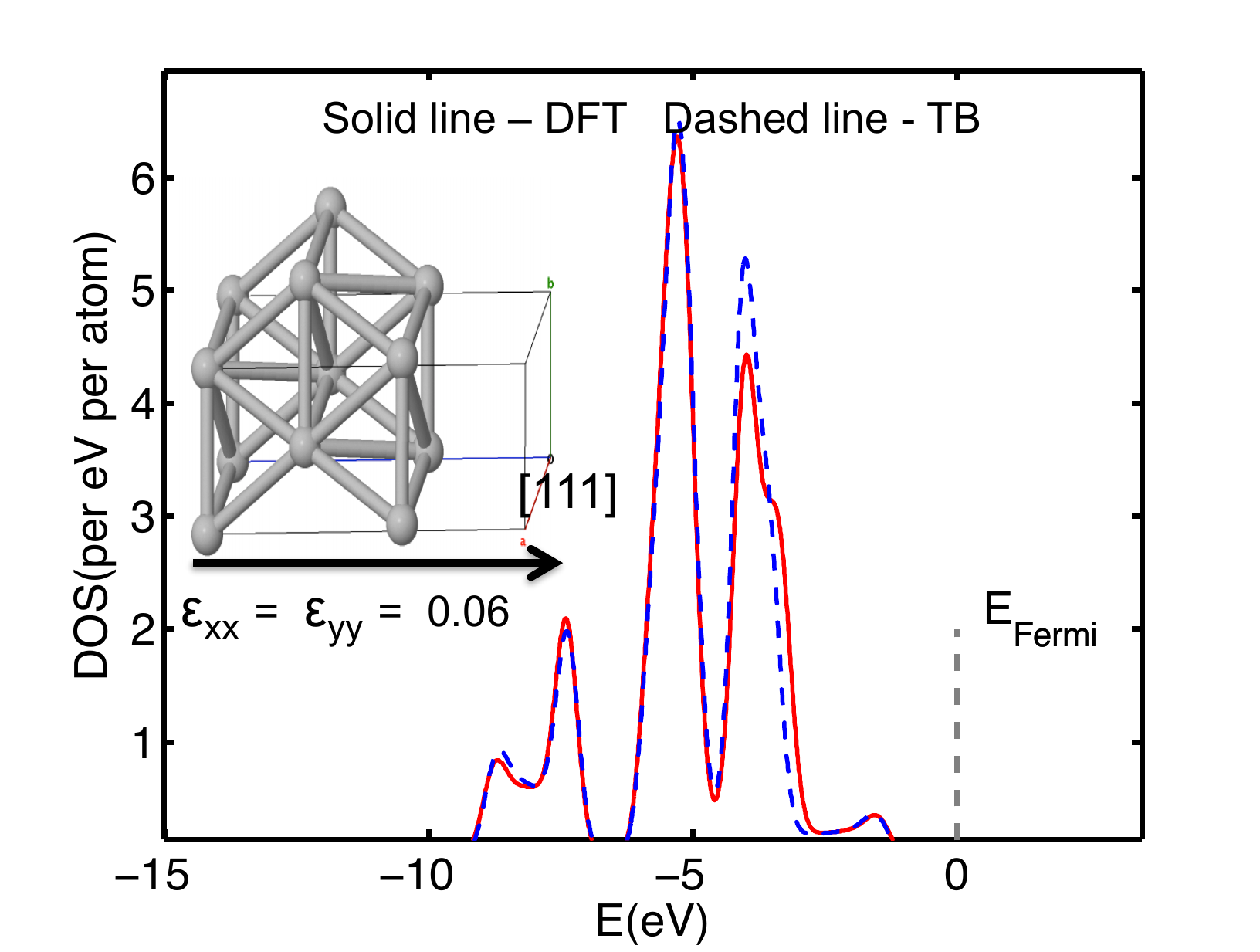}
	\caption{Unit cell structure and DOS for a Ag bulk [111] oriented unit cell strained by 6 \% in the transverse plane relative to bulk lattice parameter. Plotted here are results obtained using the TB parameters optimized in the new model (dashed lines) compared to the corresponding DFT results (solid line). The DOS is computed from the band structure for a single transverse k-point i.e. from [000] to [001] where the structure is rotated so that the z-axis coincides with the [111] direction. The band structure is normalized to $k=\pi/c$ where $c$ is the unit cell magnitude in [111]. It must be emphasized that the total DOS near the Fermi Level will not be zero when all transverse modes are included in the computation of the DOS. This result appears anomalous only because the bulk DFT band structure for $k_x=0$ and $k_y=0$ has no states close to Fermi Level.}	
	\label{Al_biax_structure_and_bs}
\end{figure}
\begin{figure}[H]
	\centering
		\includegraphics[width=3.5in]{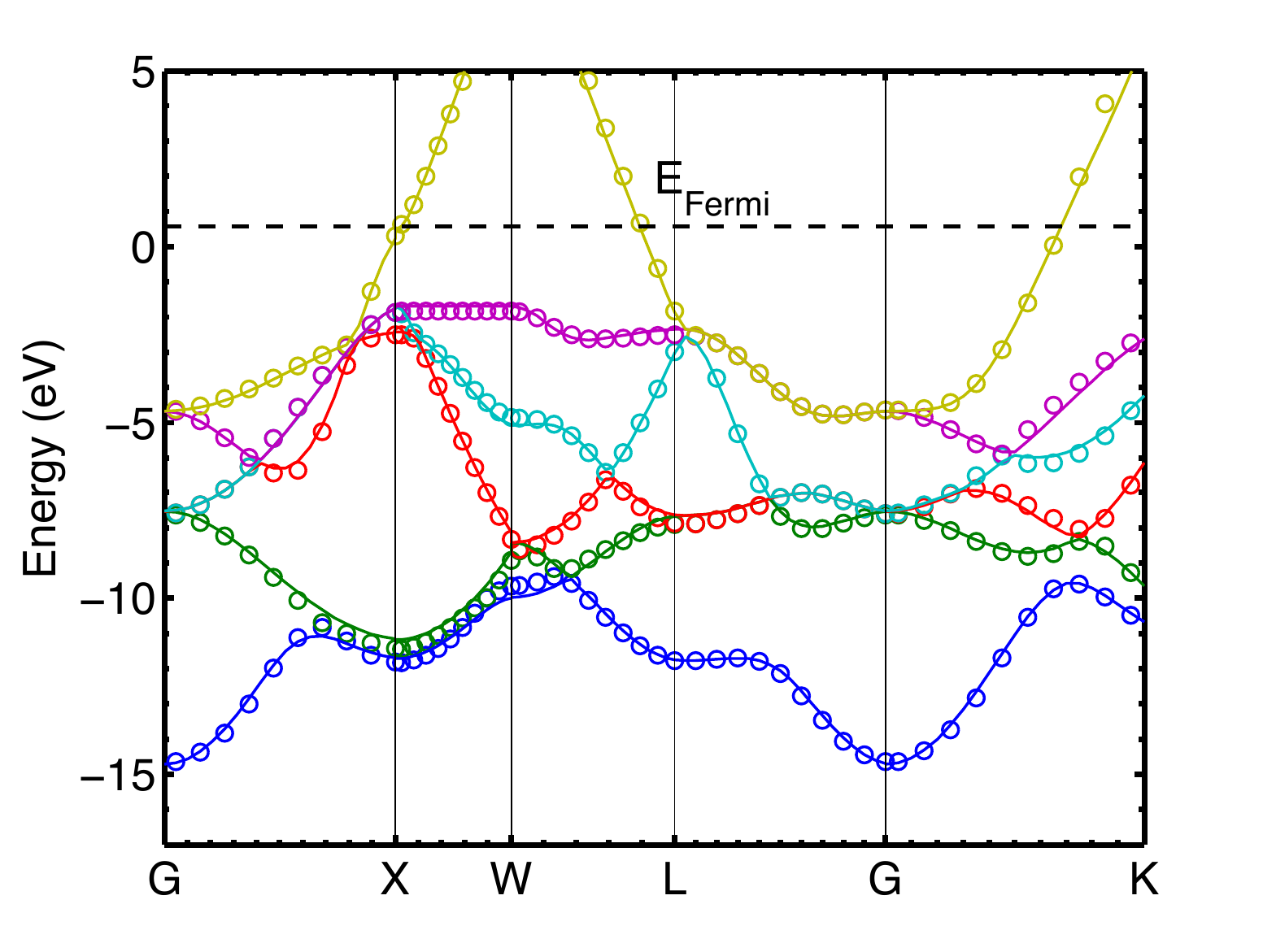}
	\caption{ Band structure of bulk Au strained homogeneously by 11 \% obtained using TB parameters optimized in the new model (circles) compared to the corresponding DFT band structure (solid lines). The dashed line indicates the Fermi Level.}	
	\label{Au_bulk_fcc_3p61_bs}
\end{figure}

\section{Alloy electronic structure for the Au-Ag system}

\begin{figure}[H]
	\centering
		\includegraphics[width=3.5in]{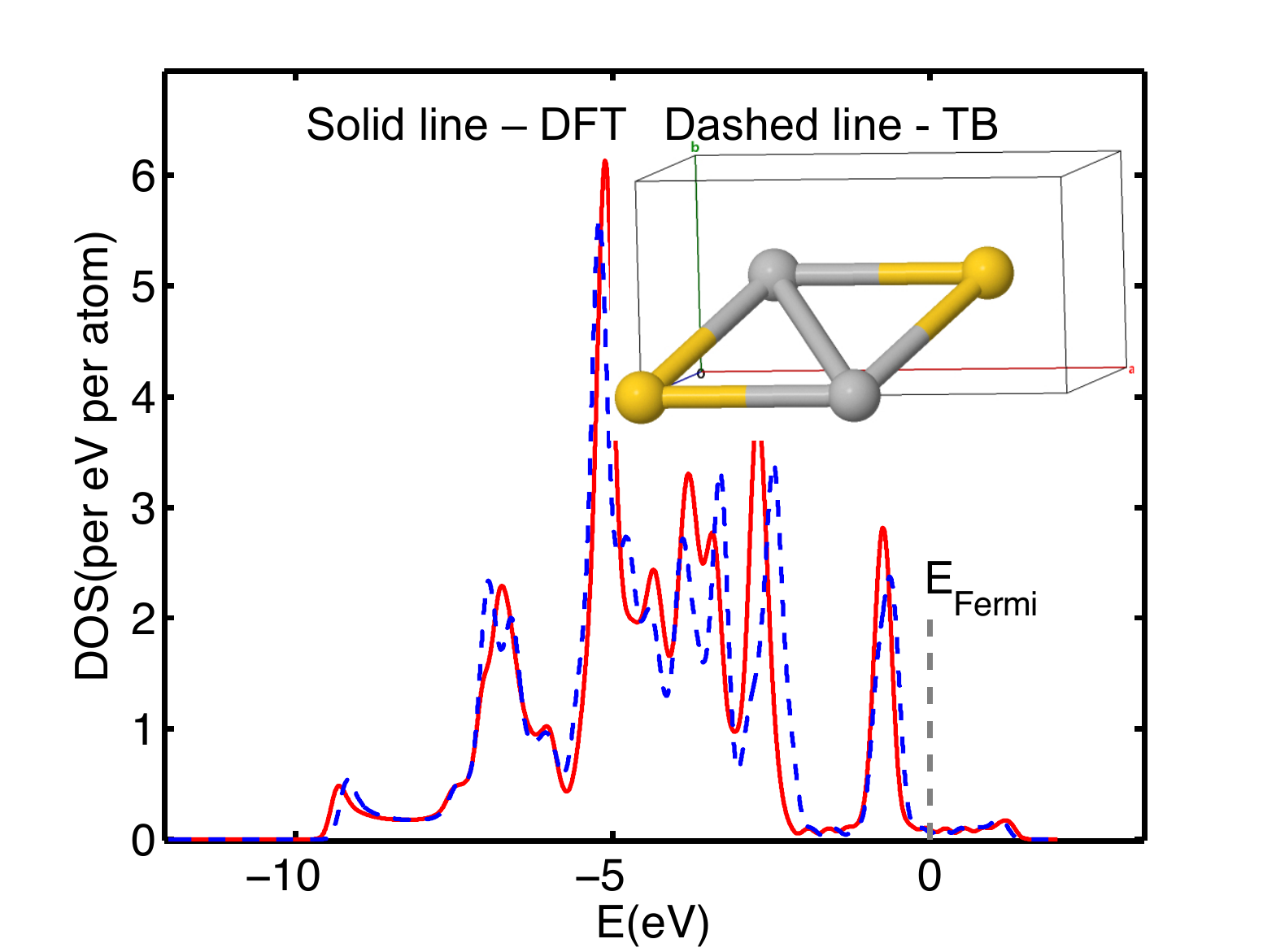}
	\caption{ Unit Cell structure and DOS of ordered 50\% Ag and 50\% Au alloy supercell obtained using TB parameters optimized in the new model (dashed line) compared to the corresponding DFT results (solid line). Silver-colored spheres represent Ag atoms while the golden spheres represent Au atoms.}	
	\label{Au50Ag50_structure_and_bs}
\end{figure}

\section{Cu Nanostructures}
\begin{figure}[H]
	\centering
		\includegraphics[width=3.5in]{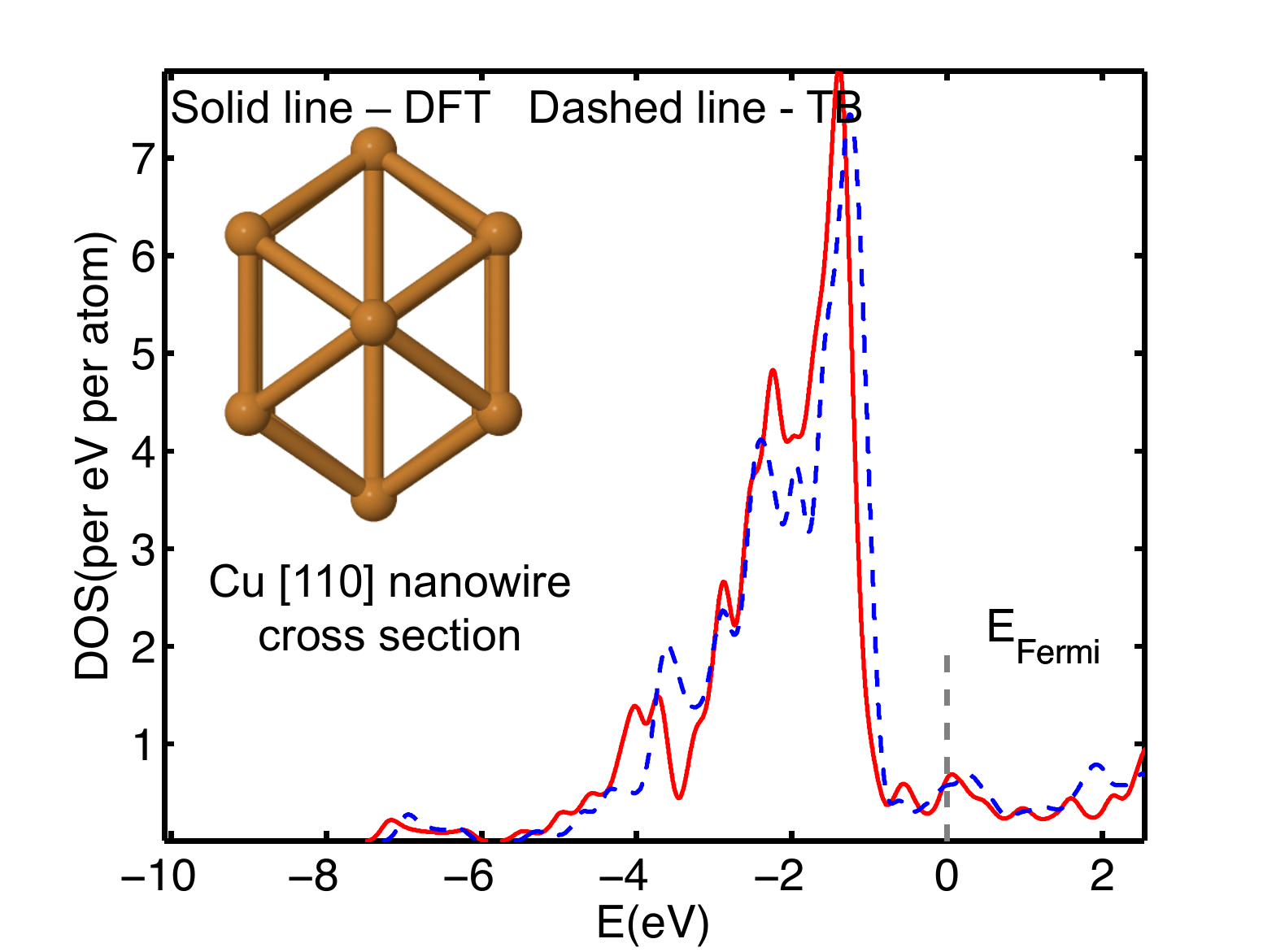}
	\caption{ Atomic structure and DOS of Cu [110] oriented nanowire, obtained using TB parameters optimized in the new model (dashed line) compared to the corresponding DFT result (solid line). The [110] direction is perpendicular to the plane of the paper and the unit cell is periodic in this direction}	
	\label{Cu_110_bs_and_structure}
\end{figure}
\begin{figure}[H]
	\centering
		\includegraphics[width=3.5in]{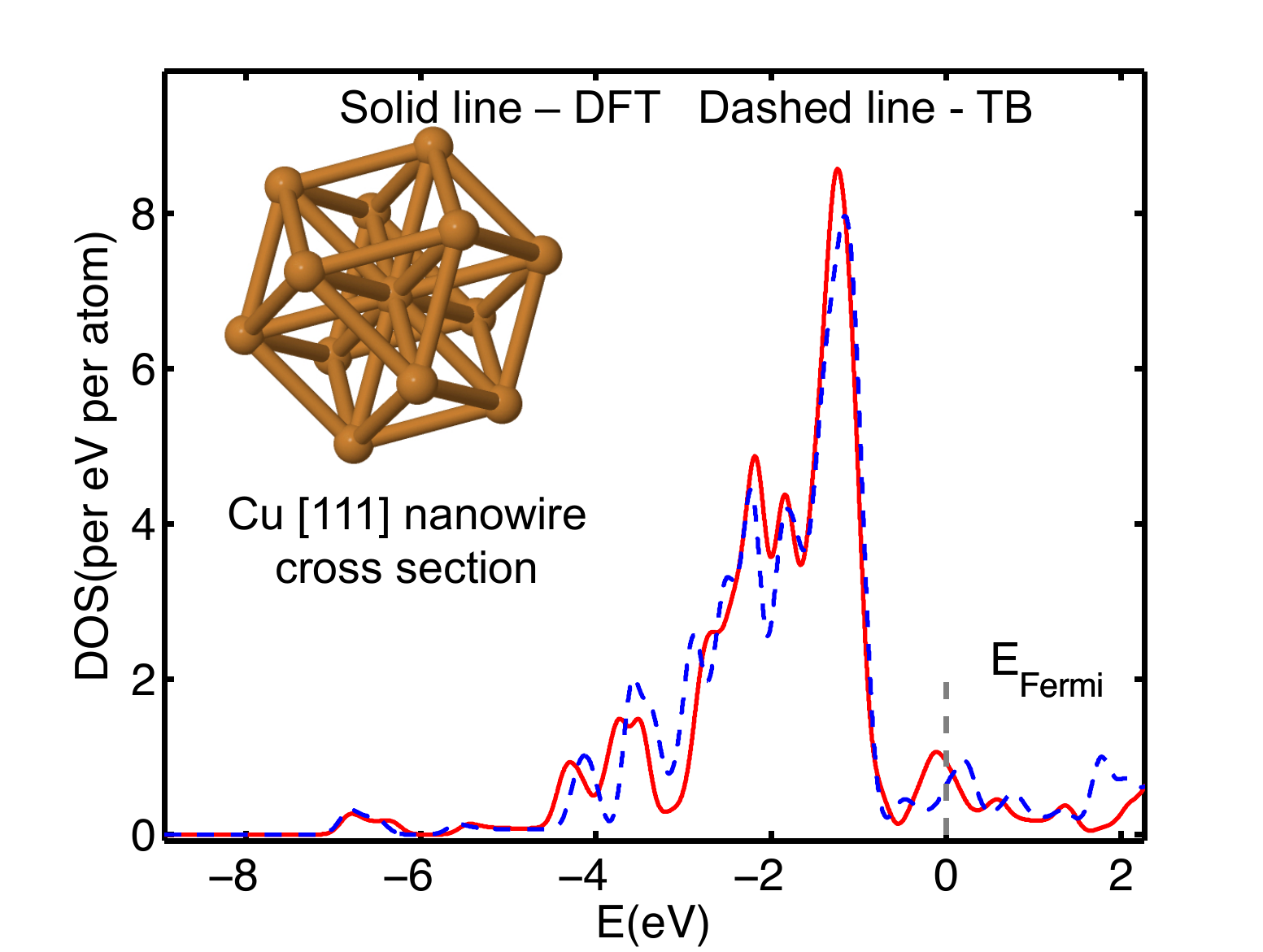}
	\caption{ Atomic structure and DOS of Cu [111] oriented nanowire, obtained using TB parameters optimized in the new model (dashed line) compared to the corresponding DFT band structure (solid line). The [111] direction is perpendicular to the plane of the paper and the unit cell is periodic in this direction}	
	\label{Cu_111_bs_and_structure}
\end{figure}

\bibliography{paper1}

\end{document}